\documentclass{aa}
\usepackage{txfonts}
\usepackage{graphicx}
\usepackage{caption}
\usepackage{subcaption}
\usepackage{color}
\usepackage{tablefootnote}
\usepackage[colorlinks=true,allcolors=blue]{hyperref}

\usepackage{tikz}
\usetikzlibrary{arrows,shapes,backgrounds}

\begin{document} 

\titlerunning{}
\authorrunning{Mountrichas et al.}
\titlerunning{}

\title{Comparison of star formation histories of AGN and non-AGN galaxies}

\author{G. Mountrichas\inst{1}, V. Buat\inst{2,3}, G. Yang\inst{4,5}, M. Boquien\inst{6}, Q. Ni\inst{7}, E. Pouliasis\inst{8}, D. Burgarella\inst{2}, P. Theule\inst{2}, I. Georgantopoulos\inst{8}}
          
     \institute {Instituto de Fisica de Cantabria (CSIC-Universidad de Cantabria), Avenida de los Castros, 39005 Santander, Spain
              \email{gmountrichas@gmail.com}
              \and
             Aix Marseille Univ, CNRS, CNES, LAM Marseille, France. 
              \and
                 Institut Universitaire de France (IUF)
                 	\and
                Department of Physics and Astronomy, Texas A\&M University, College Station, TX 77843-4242, USA 
                \and
                George P. and Cynthia Woods Mitchell Institute for Fundamental Physics and Astronomy, Texas A\&M University, College Station, TX 77843-4242, USA 
                	\and
                Centro de Astronom\'ia (CITEVA), Universidad de Antofagasta, Avenida Angamos 601, Antofagasta, Chile
			\and
			Institute for Astronomy, University of Edinburgh, Royal Observatory, Edinburgh, EH9 3HJ, UK
			 \and
             National Observatory of Athens, Institute for Astronomy, Astrophysics, Space Applications and Remote Sensing, Ioannou Metaxa and Vasileos Pavlou GR-15236, Athens, Greece}

\abstract {We use AGN with X-ray luminosities, $\rm L_{X,2-10keV} \sim 10^{42.5-44}\,erg\,s^{-1}$, from the {\it{COSMOS-Legacy}} survey that lie within the UltraVISTA region and cross match them with the LEGA-C catalogue. The latter provides measurements of the calcium break, D$_n$4000, and H$_\delta$ Balmer line that allow us to study the stellar populations of AGN and compare them with a galaxy reference catalogue. Our samples consist of 69 AGN and 2176 non-AGN systems, within $\rm 0.6<z<1.3$, that satisfy the same photometric selection criteria. We construct the SEDs of both population and use the CIGALE code to investigate the effect of the two indices in the SED fitting process. Our analysis shows that the inclusion of D$_n$4000 and H$_\delta$ allows CIGALE to constrain better the ages of the stellar populations. Furthermore, we find an increase of the estimated stellar masses by, on average, $\sim 0.2$\,dex, in particular for systems with young stars (D$_n$4000$\,<1.5$), when the two indices are included in the SED fitting. We then compare the D$_n$4000 and H$_\delta$ of AGN with sources in the reference catalogue, accounting for the different stellar mass of the two populations. Our analysis reveals that low to moderate L$_X$ AGN tend to reside in galaxies with older stellar populations and are less likely to have experienced a recent star formation burst, compared to galaxies in the control sample. Finally, we compare the two populations as a function of their morphology (bulge-dominated, BD, vs. non-BD) and compactness (mass-to-size ratio). A similar fraction of AGN and non-AGN systems are classified as non-BD ($\sim 70\%$). Our analysis shows that BD AGN tend to have younger stellar populations compared to BD, non-AGN systems. On the other hand, non-BD AGN have, on average, older stellar populations and are less likely to have experienced a burst compared to non-BD sources in the reference sample. Furthermore, AGN tend to prefer more compact systems compared to non-AGN.}


\keywords{}
   
\maketitle

\section{Introduction}

It is widely accepted and well established that there is a correlation between the mass of a supermassive black hole (SMBH) and the properties of the galactic bulge \citep[e.g.,][]{Magorrian1998, Ferrarese2000}. It is also known that black holes grow through accretion of cold gas onto their accretion disc. When this happens the SMBH becomes active and the galaxy is called an active galactic nuclei (AGN). The gas that triggers the SMBH originates either from the host galaxy or the extragalactic environment. Various mechanisms have been suggested to drive the gas over more than nine orders of magnitude, from kiloparsec to sub-parsec scales \citep[e.g.,][]{Alexander2012}. However, the cold gas does not only activate the SMBH, but it can also trigger the star formation of the galaxy. Furthermore, a number of studies have found that AGN activity and star formation peak at the same cosmic time \citep[$\rm z\sim2$; e.g.,][]{Boyle1998, Boyle2000, Sobral2013}. These advocate for a connection between the BH activity and galaxy growth. However, the nature of this connection is still elusive.

Hydrodynamical simulations and semi-analytic models have shown that AGN feedback, in the form of, for example, jets, winds and outflows can affect the star formation of the host galaxy both ways \citep[e.g.,][]{Zubovas2013}. It can either quench the formation of stars \citep[e.g.,][]{DeBuhr2012} or enhance it \citep[e.g.,][]{Zinn2013}. Results from observational studies that have compared the SFR of AGN with that of star forming main sequence (SFMS) galaxies suggest that low to moderate X-ray luminosity AGN ($\rm L_{X,2-10keV} < 10^{44}\,erg\,s^{-1}$) present lower or consistent SFR with that of SFMS galaxies, while the most luminous AGN have enhanced SFR compared to SFMS galaxies \citep[e.g.,][]{Santini2012, Shimizu2015, Shimizu2017, Masoura2018, Florez2020, Mountrichas2021a, Mountrichas2022a, Mountrichas2022b, Pouliasis2022}. This suggests different star formation histories (SFH) for AGN and SFMS galaxies, i.e., differences in the timescales that the stars are formed and the mechanisms (e.g., mergers) that govern the star formation of each population.

The Large Early Galaxy Astrophysics Census (LEGA-C) survey \citep{Wel2016, Straatman2018} has collected high signal-to-noise, high resolution spectra for $\sim 3500$ galaxies that lie within $\rm 0.6<z<1.3$ . This allows the study of the ages and metallicities of the stellar populations of these galaxies as well as their stellar kinematics. \cite{Wu2018}, used measurements provided in the LEGA-C catalogue for two age sensitive absorption line indices, the equivalent width (EW) of H$_\delta$ Balmer line \citep{Worthey_ottaviani1997} and the calcium break, D$_n$4000 \citep{Balogh1999}, to study the stellar populations of the galaxies included in the LEGA-C dataset. D$_n$4000 is small for young stellar populations and large for old, metal rich galaxies. On the other hand, the EW of H$_\delta$ rises rapidly in the first few hundred Myrs after a burst of star formation, when O- and B-type stars dominate the spectrum and then decreases when A-type stars fade \citep[e.g.,][]{Kauffmann2003a, Wu2018}. Based on the analysis of \cite{Wu2018}, galaxies at $\rm z\sim 0.8$ present a bimodal D$_n$4000-H$_\delta$ distribution, implying a bimodal light-weighted age distribution. \cite{Sobral2022}, used the LEGA-C catalogue and the two indices (D$_n$4000, H$_\delta$) to study the stellar populations of central and satellites galaxies. Their analysis showed that for star forming galaxies, D$_n$4000 and H$_\delta$ depend on stellar mass, M$_*$, and are completely independent of the environment.

The two spectral indices have also been used to study the stellar populations of AGN. \cite{Kauffmann2003b} examined the stellar populations of $\sim 22500$ narrow line AGN, at $\rm 0.02<z<0.3$, selected from the Sloan Digital Sky Survey \citep[SDSS;][]{York2000, Stoughton2002}. They found that the host galaxies of low luminosity AGN $(\rm log\,L[OIII]<7\,L_\sun)$ have stellar populations similar to normal early-types, while the high luminosity AGN have much younger stellar ages. A significant fraction of the more luminous AGN have strong H$_\delta$ lines. They also examined a sample of broad line AGN and found no significant differences between broad and narrow line AGN. Recently, Georgantopoulos et al. (submitted) compared the ages and galaxy properties of 55 X-ray unobscured or mildly obscured AGN ($\rm N_H<10^{23}\,cm^{-2}$) and 25 heavily obscured sources ($\rm N_H>10^{23}\,cm^{-2}$) in the COSMOS field, at $\rm 0.6<z<1.0$. They found that the majority of unobscured AGN appear to live in younger galaxies in contrast to the obscured AGN which tend to reside in systems with older ages located between the young and old galaxy population.

In this work, we use low to moderate X-ray luminosity ($\rm L_{X,2-10keV} \sim 10^{42.5-44}\,erg\,s^{-1}$) AGN from the UltraVISTA region of the {\it{COSMOS-Legacy}} survey \citep{Marchesi2016, Civano2016} and cross correlate them with the LEGA-C catalogue. This enriches our sample with the D$_n$4000 and H$_\delta$ measurements included in the LEGA-C dataset. We also compile a galaxy reference catalogue (control sample) that lies within the same area and redshift range as the AGN ($\rm 0.6<z<1.3$) for which we also have available D$_n$4000 and H$_\delta$ measurements from the LEGA-C catalogue (Sect. 2). First, we construct the spectral energy distributions (SEDs) of both AGN and sources in the reference galaxy catalogue, using photometry from optical to far infrared and fit their SEDs using the CIGALE code \citep{Boquien2019, Yang2020, Yang2022}. The purpose of this exercise is to examine the effect of the inclusion of D$_n$4000 and H$_\delta$ in the SED fitting measurements (Sect. 3). Specifically, we want to check whether the two spectral indices help CIGALE break the degeneracies of the SFH parameters, leading to more robust measurements of the galaxy properties. Then, we compare the D$_n$4000 and H$_\delta$  of AGN with those of sources in the reference catalogue (sections 4.1, 4.2, 4.3). Moreover, we examine whether the morphology and compactness of the (host) galaxy affects this comparison, using the catalogue presented in \cite{Ni2021} (Sect. 4.4).


Throughout this work, we assume a flat $\Lambda$CDM cosmology with $H_ 0=70.4$\,Km\,s$^{-1}$\,Mpc$^{-1}$ and $\Omega _ M=0.272$ \citep{Komatsu2011}.

 

\begin{figure}
\centering
  \includegraphics[width=0.8\linewidth, height=6.5cm]{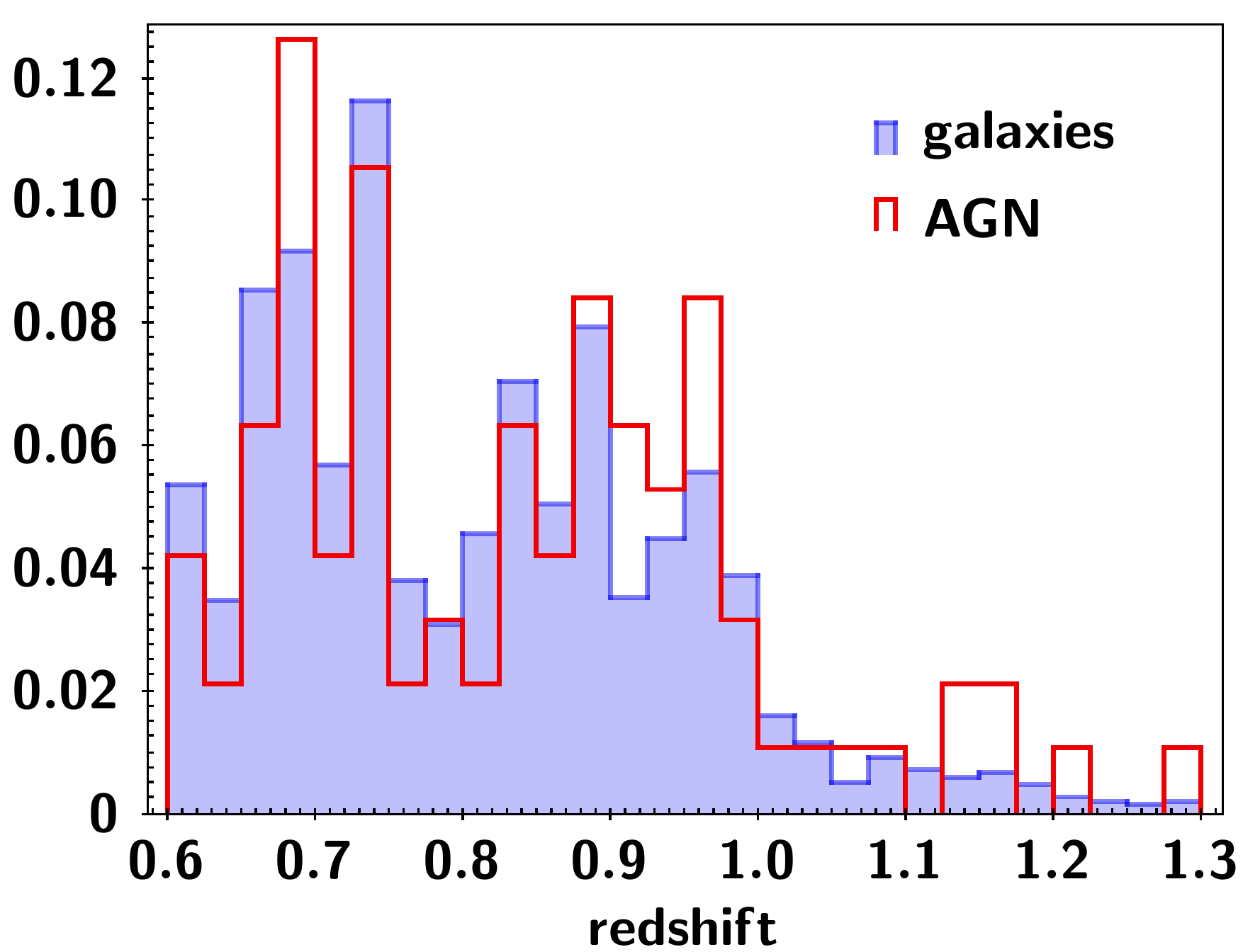}
  \caption{Redshift distributions of sources in the galaxy reference catalogue (blue shaded histogram) and of X-ray AGN (red line). The two populations present similar distributions.}
  \label{fig_redz}
\end{figure} 


\begin{figure}
\centering
  \includegraphics[width=0.8\linewidth, height=6.5cm]{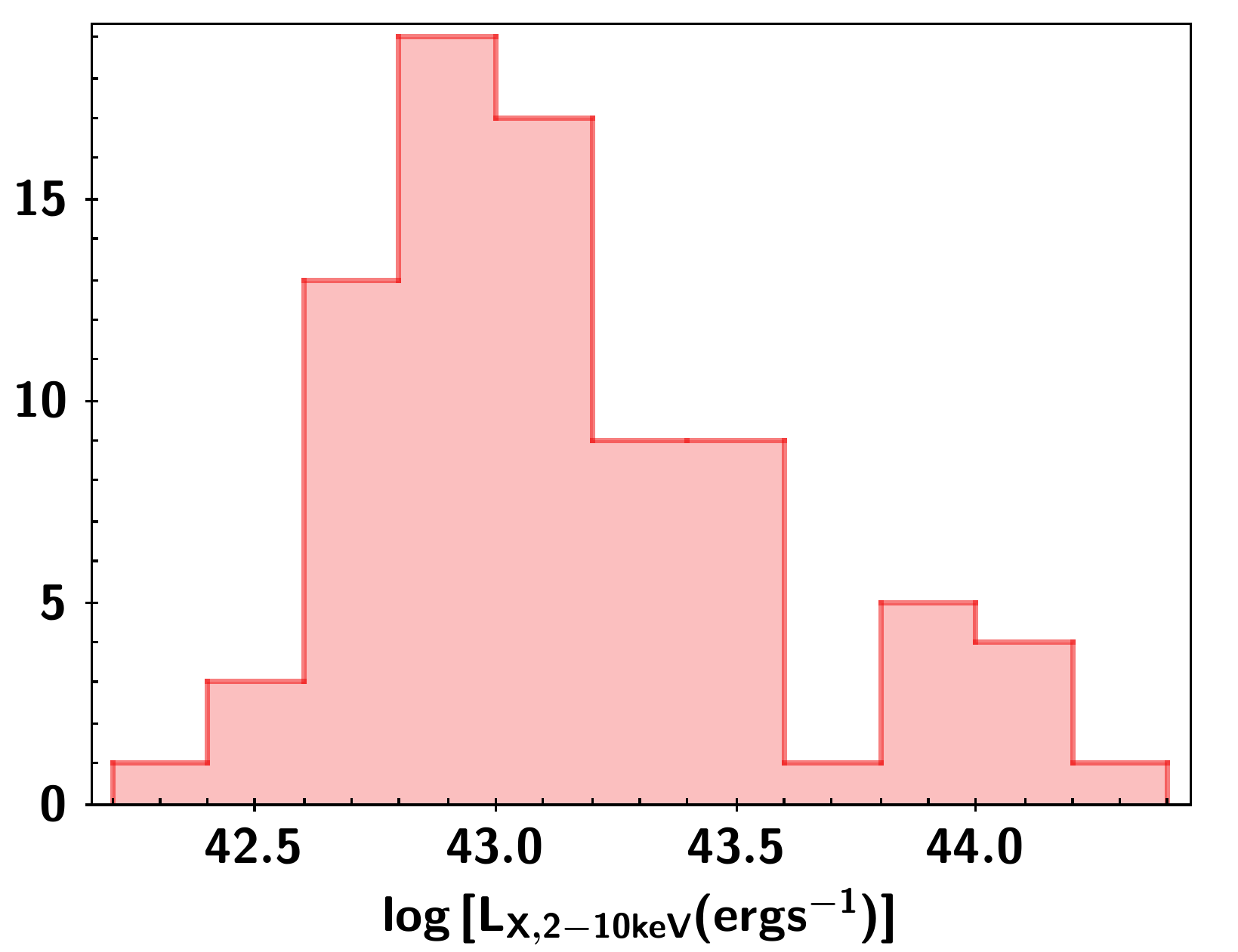}
  \caption{X-ray luminosity distributions in the $2-10$\,keV band of the X-ray AGN used in our analysis. The vast majority of sources have low to moderate X-ray luminosities ($\rm L_{X,2-10keV} \leq 10^{44}\,erg\,s^{-1}$).}
  \label{fig_lx}
\end{figure}

\section{Data}
\label{sec_data}

\subsection{The LEGA-C catalogue}
\label{sec_legac_catalogue}

The LEGA-C catalogue includes data obtained from the LEGA-C survey \citep{Wel2016, Straatman2018}. In this work, we use the third and final data release of the catalogue \citep{Wel2021} that includes 4081 galaxy spectra, with 3741 unique objects that cover a redshift range from 0.6 to 1.3. The survey is based on a parent sample of spectroscopic and photometric galaxies from the UltraVISTA region \citep{Muzzin2013} of the COSMOS field \citep{Scoville2007} and covers a footprint of 1.4255 square degrees. The key characteristic of the survey is that it is $K_S-$band selected. Specifically, the targeted sources are brighter than a redshift dependent $K_S-$ limit ($K_{S,LIM}=20.7-7.5\,\rm{log(1+z)}/1.8$).

The catalogue includes high resolution and signal-to-noise (SNR) spectra ($R \sim 3500, \rm typical\,SNR\approx 20\AA ^{-1}$) that have allowed the measurement of stellar velocity dispersions, stellar population properties and absorption line indices, as well as, emission line fluxes and equivalent widths, among others. In this work, we make use of two stellar age sensitive tracers, the equivalent width of H$\delta$ absorption and D$_n$4000 index \citep[e.g.,][]{Kauffmann2003, Wu2018}. Comparison of their values between X-ray selected AGN and a reference catalogue of galaxy (non-AGN) sources will allow us to draw conclusions about the star formation history of the two populations.  

To measure the two indices the stellar continuum is separated from the ionized gas emission and the observed spectrum is modelled using the Penalized Pixel-Fitting (pPXF) method \cite{Cappellari2004}. Galaxy spectrums are fit by a combination of stellar and gas emission templates. For sources included in our sample the average uncertainty on H$\delta$ and D$_n$4000 is 17\% and 3\%, respectively. This is true for the AGN and sources in the galaxy reference catalogue (see next Section). A detailed description of the process is provided in \cite{Bezanson2018} and \cite{Wu2018}. For D$_n$4000 and H$_\delta$ the definitions in \cite{Balogh1999} and \cite{Worthey_ottaviani1997} have been used, respectively.

\subsection{X-ray AGN and galaxy reference catalogues}

The X-ray and galaxy reference catalogues used in this work, have been described in detail in Sect. 2 in \cite{Mountrichas2022a}. Below there is a brief description of the two datasets. First, we present the available photometry of the samples and then we add the available information on the spectral indices and describe the properties of the final catalogues.

\subsubsection{X-ray sample}
The X-ray sample has been extracted from the X-ray dataset described in \cite{Marchesi2016} and includes observations from the {\it{COSMOS-Legacy}} survey \citep{Civano2016}. The latter is a 4.6\,Ms {\it{Chandra}} program that covers 2.2\,deg$^2$ of the COSMOS field \citep{Scoville2007}. The X-ray catalogue includes 4016 sources. \cite{Marchesi2016} matched the X-ray sources with optical and infrared counterparts using the likelihood ratio technique \citep{Sutherland_and_Saunders1992}. Of the sources, 97\%  have an optical and IR counterpart and a photometric redshift and $\approx 54\%$  have spectroscopic redshift. Only X-ray sources within both the COSMOS and UltraVISTA \citep{McCracken2012} regions are used in our analysis. The reason for restricting the X-ray dataset in the UltraVISTA region is that the LEGA-C catalogue with which we cross-match the X-ray sample consists of galaxies in this region. This reduces the number of AGN to 1718 X-ray detected sources with $\rm log\,[L_{X,2-10keV}(ergs^{-1})]>42$.

A subsidiary goal of this work is to examine the effect of the inclusion of the two spectral lines (H$\delta$, D$_n$4000) in the measurements of the star formation history (SFH) parameters and (host) galaxy properties estimated by SED fits (see next section). To construct the SEDs, the X-ray catalogue is cross-matched with the COSMOS photometric catalogue produced by the {\it{Herschel}} Extragalactic Legacy Project (HELP) collaboration \citep{Shirley2019, Shirley2021}. HELP includes data from 23 of the premier extragalactic survey fields imaged by the {\it{Herschel}} Space Observatory which form HELP. The catalogue provides homogeneous and calibrated multiwavelength data. The cross-match with the HELP catalogue is done using a 1\arcsec radius and the optical coordinates of the counterpart of each X-ray source. To obtain reliable measurements through the SED fitting process, we require all our X-ray AGN to have been detected in the following photometric bands $u, g, r, i, z, J, H, K_s$, IRAC1, IRAC2, and MIPS/24. IRAC1, IRAC2, and MIPS/24 are the [3.6]\,$\mu$m, [4.5]\,$\mu$m, and 24 $\mu$m photometric bands of Spitzer \citep{Mountrichas2022b}. This photometric criterion reduces the X-ray sample to 1627 AGN. All these sources have measured fluxes in the  {\it{Herschel}} PACS photometric bands and $\sim 80\%$ also have SPIRE bands. \cite{Mountrichas2022a} examined the effect of the lack of far-IR photometry on SFR measurements, by applying SED fitting with and without far-IR photometry, on sources in the COSMOS field  (see their Sect. 3.2.2). Based on their results, the mean difference, $\mu$, of the SFR calculations is 0.01 and the dispersion, $\sigma$, is 0.25. Similar numbers are found for sources in the galaxy reference catalogue (see below; $\mu=0.05$ and $\sigma=0.16$).

\begin{table*}
\caption{Models and the values for their free parameters used by X-CIGALE for the SED fitting.} 
\centering
\setlength{\tabcolsep}{1.mm}
\begin{tabular}{cc}
       \hline
Parameter &  Model/values \\
        \hline
\multicolumn{2}{c}{Star formation history: delayed model and recent burst} \\
Age of the main population & 3000, 4000, 5000, 5500, 6000 Myr \\
e-folding time & 50, 100, 200, 500, 700 Myr \\ 
Age of the burst & 50 Myr \\
Burst stellar mass fraction & 0.0, 0.001, 0.003, 0.005, 0.01, 0.02, 0.1   \\
\hline
\multicolumn{2}{c}{Simple Stellar population: Bruzual \& Charlot (2003)} \\
Initial Mass Function & Chabrier (2003)\\
Metallicity & 0.02 (Solar) \\
\hline
\multicolumn{2}{c}{Galactic dust extinction: dustatt\_modified\_CF00} \\
Dust attenuation law & Charlot \& Fall (2000) law   \\
V-band attenuation $A_V$ & 0.2, 0.3, 0.4, 0.5, 0.6, 0.7, 0.8, 0.9, 1, 1.5, 2, 2.5, 3, 3.5, 4 \\ 
\hline
\multicolumn{2}{c}{Galactic dust emission: Dale et al. (2014)} \\
$\alpha$ slope in $dM_{dust}\propto U^{-\alpha}dU$ & 2.0 \\
\hline
\multicolumn{2}{c}{AGN module: SKIRTOR} \\
Torus optical depth at 9.7 microns $\tau _{9.7}$ & 3.0, 7.0 \\
Torus density radial parameter p ($\rho \propto r^{-p}e^{-q|cos(\theta)|}$) & 1.0 \\
Torus density angular parameter q ($\rho \propto r^{-p}e^{-q|cos(\theta)|}$) & 1.0 \\
Angle between the equatorial plan and edge of the torus & $40^{\circ}$ \\
Ratio of the maximum to minimum radii of the torus & 20 \\
Viewing angle  & $30^{\circ}\,\,\rm{(type\,\,1)},70^{\circ}\,\,\rm{(type\,\,2)}$ \\
AGN fraction & 0.0, 0.1, 0.2, 0.3, 0.4, 0.5, 0.6, 0.7, 0.8, 0.9, 0.99 \\
Extinction law of polar dust & SMC \\
$E(B-V)$ of polar dust & 0.0, 0.2, 0.4 \\
Temperature of polar dust (K) & 100 \\
Emissivity of polar dust & 1.6 \\
\hline
\multicolumn{2}{c}{X-ray module} \\
AGN photon index $\Gamma$ & 1.4 \\
Maximum deviation from the $\alpha _{ox}-L_{2500 \AA}$ relation & 0.2 \\
\hline
Total number of models & 24,948,000 (3,118,500 per redshift)   \\
\hline
\label{table_cigale}
\end{tabular}
\tablefoot{For the definition of the various parameters, see section \ref{sec_cigale}.}
\end{table*}

\subsubsection{Galaxy reference catalogue} 

In our analysis, we compare the H$\delta$ and D$_n$4000 of X-ray AGN with that of non-AGN galaxies. The galaxy reference catalogue is provided by the HELP collaboration. About $500\,000$ sources are in the UltraVISTA region and approximately $230\,000$ meet the photometric criteria applied on the X-ray dataset. About 50\% of the sources in the galaxy reference catalogue also have available measurements in the far-IR ({\it{Herschel}}).

\subsubsection{Final LEGA-C samples}

We cross match the X-ray and galaxy reference catalogue described above with the LEGA-C dataset, using a 1\arcsec radius and the optical coordinates provided in each catalogue. 134 X-ray AGN and 3,105 sources in the reference catalogue have counterparts in the LEGA-C sample. We exclude sources with $\rm FLAG\_SPEC=2$, that indicates that the photometry based flux calibration showed significant imperfections, compromising the measurement of absorption and emission indices \citep[sect. 4.2 in][]{Wel2021}. We further exclude sources with $\rm FLAG\_MORPH=1$ or 2. This flag is used to indicate cases where the light coming from the slit is not from a single galaxy with a regular morphology.  Finally, we use line indices only for galaxies that have a measured stellar velocity dispersion \citep[sect. 3.4 in][]{Wel2021}. These criteria reduce the number of X-ray AGN to 94 and the number of galaxies to 2834, within $\rm 0.6<z<1.3$.


The redshift distributions of the two catalogues are presented in Fig. \ref{fig_redz}. The two populations present similar distributions. Moreover, the vast majority of the sources ($91\%$ of AGN and $93\%$ of sources in the reference catalogue) are within $\rm 0.6<z<1.0$. The small redshift range probed by the samples allows us to assume that there is no (significant) redshift evolution that could affect our results. Fig. \ref{fig_lx} shows the X-ray luminosity distribution of the AGN. The vast majority of X-ray sources ($\sim 95\%$) have low to moderate luminosities ($\rm L_{X,2-10keV} \leq 10^{44}\,erg\,s^{-1}$).


\begin{figure}
\centering
  \includegraphics[width=1.\columnwidth, height=6.5cm]{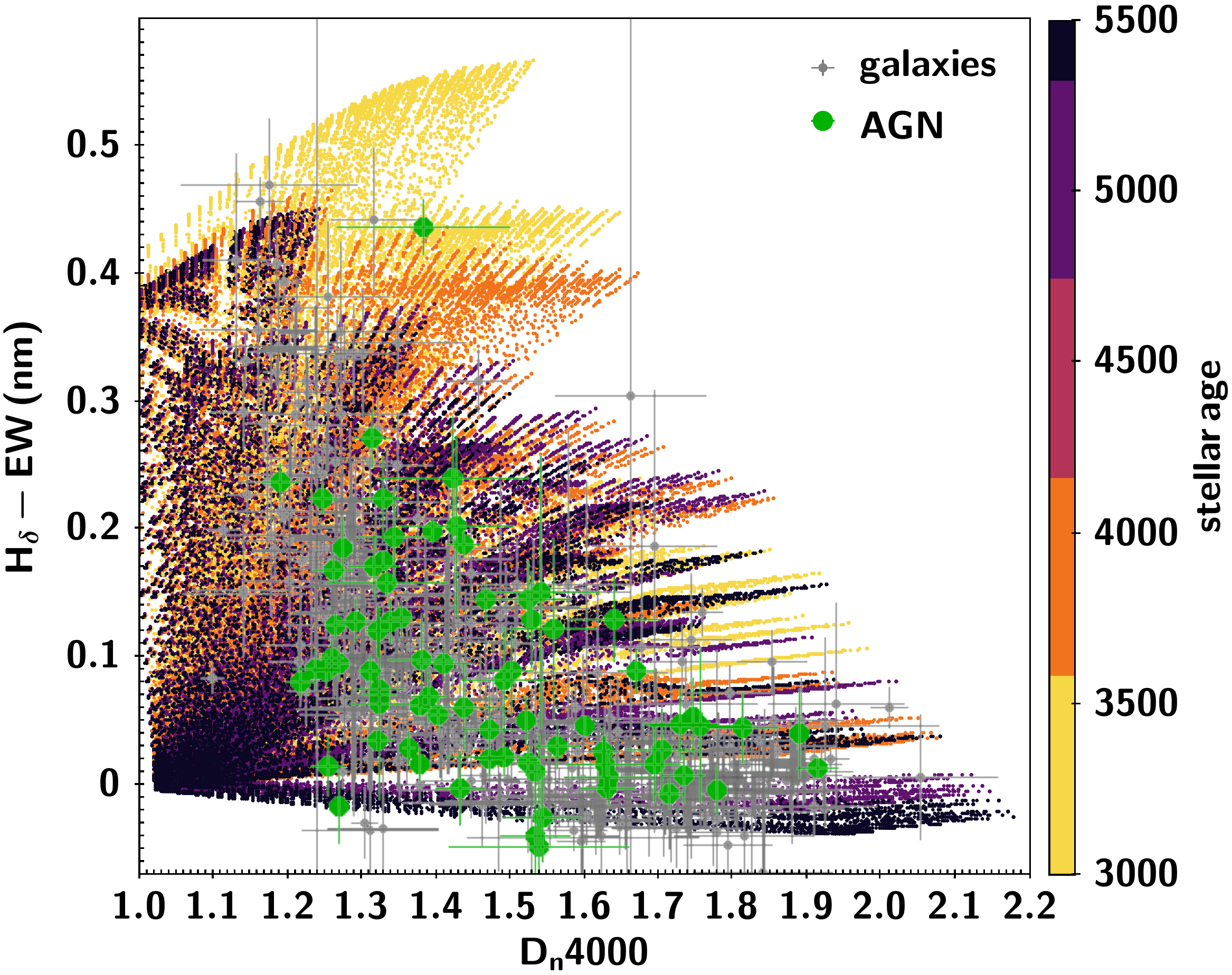} 
  \caption{The equivalent width (EW) of H$\delta$ against D$_n$4000 for the models and the AGN and sources in the reference catalogue. Models are colour coded based on the age of the stellar population, calculated by CIGALE. The parameter space we use for the SED fitting covers well the data space. The number of models and galaxies are reduced by a factor of 10 and 3, respectively, on the figure for better visibility.} 
  \label{models_vs_data}
\end{figure}


\begin{figure}
\centering
  \includegraphics[width=0.75\columnwidth, height=5.8cm]{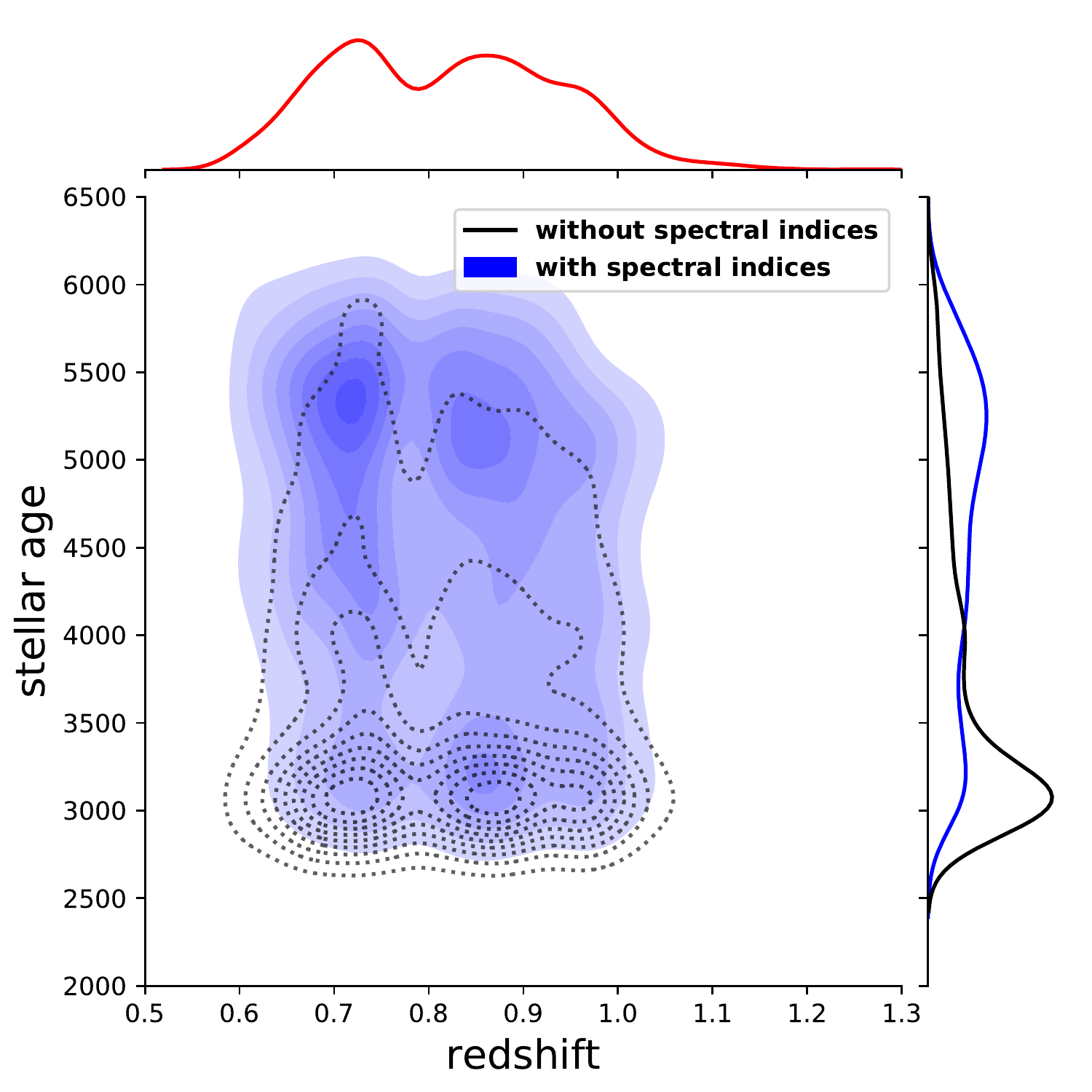} 
  \includegraphics[width=0.75\columnwidth, height=5.8cm]{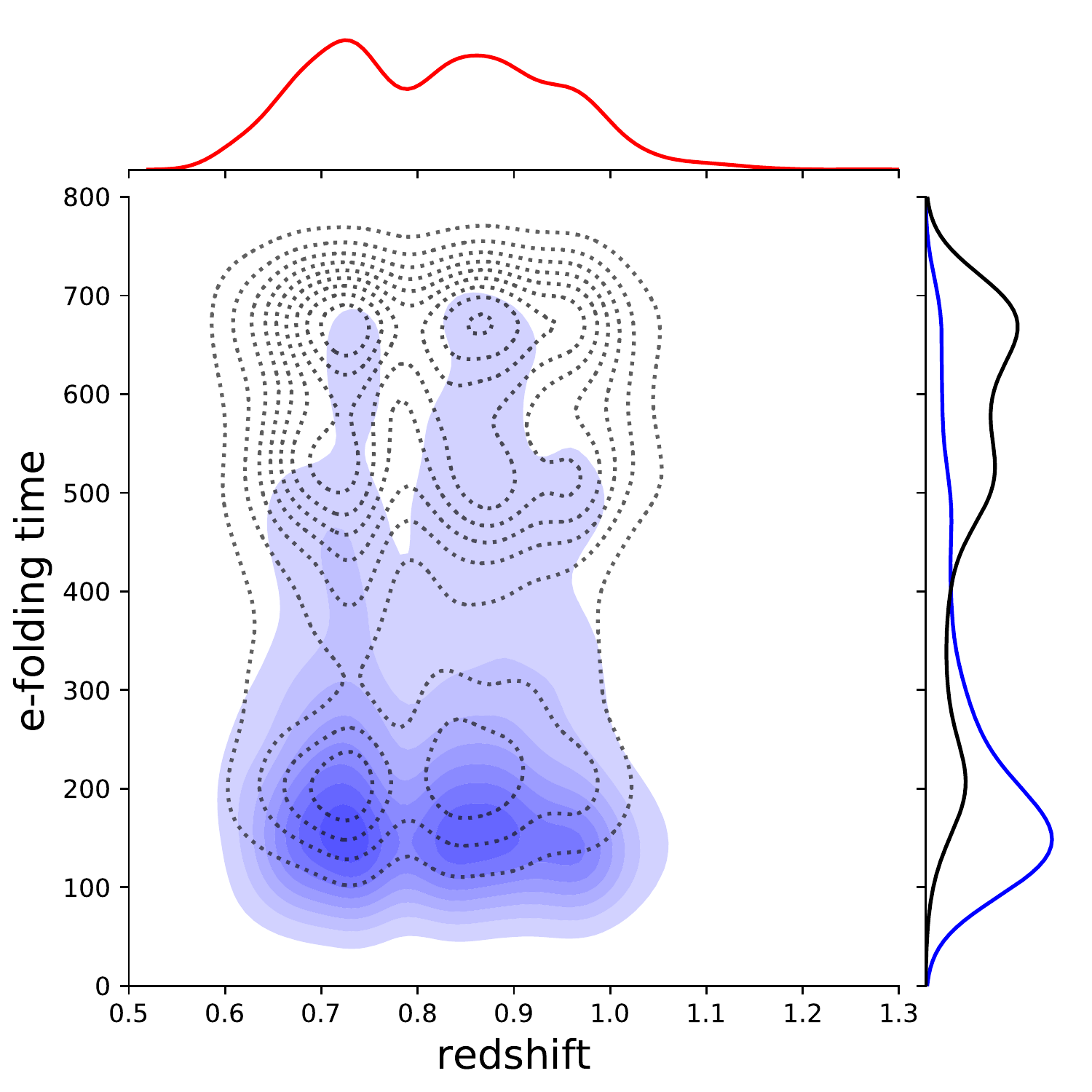} 
  \includegraphics[width=0.75\columnwidth, height=5.8cm]{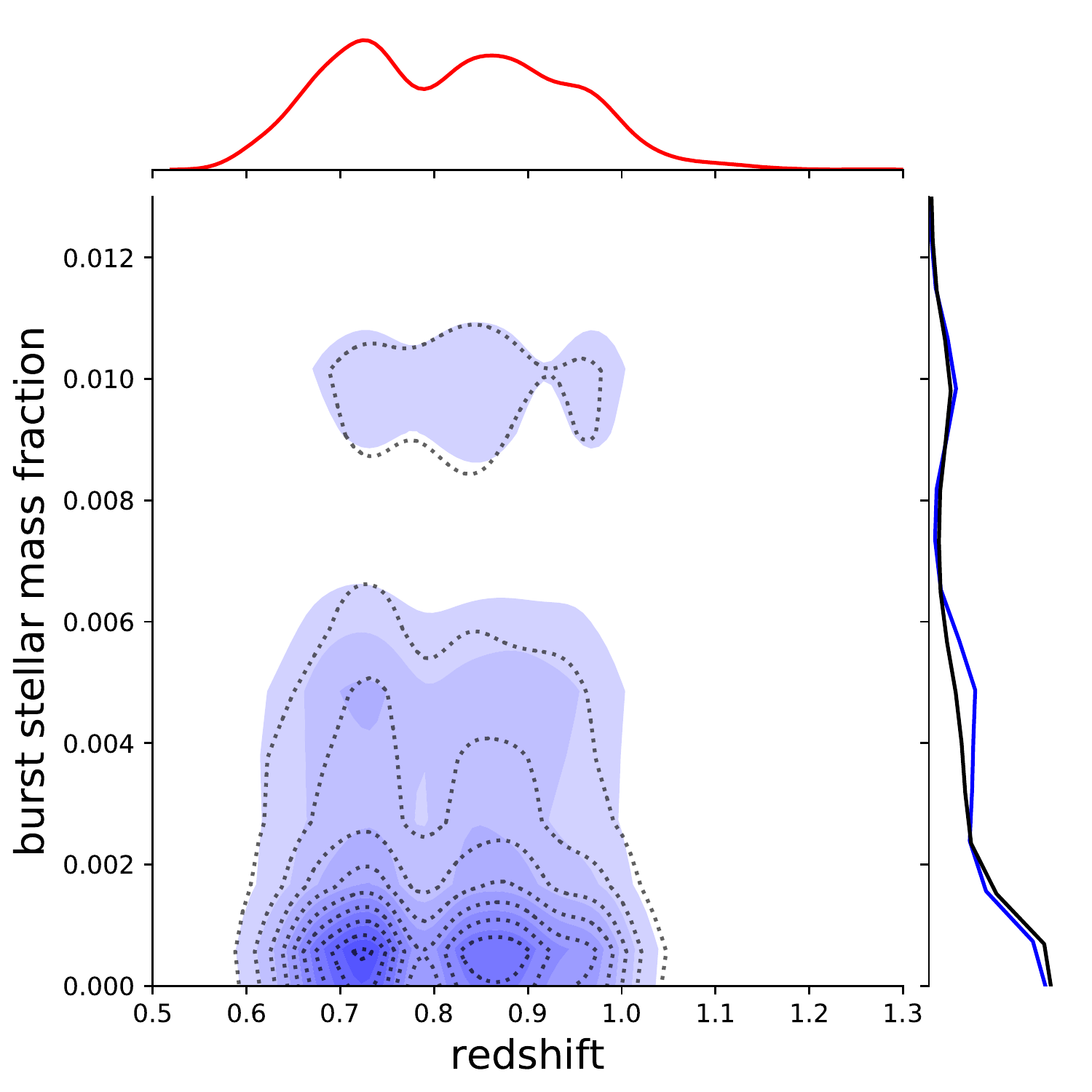}
  \caption{SFH parameters as a function of redshift, with (blue shaded contours) and without (black line contours) including the H$\delta$ and D$_n$4000 indices in the SED fitting, for sources in the galaxy reference catalogue. Top panel presents the stellar age in Myr. When the two indices are not included in the fitting process, the distribution highly peaks at the lowest stellar age value, allowed by the parametric grid (see Table \ref{table_cigale}). This is at odds with the redshift distribution of the galaxies. When the two indices are included in the fitting process, the distribution of stellar ages appears flatter and similar to the redshift distribution. The middle panel presents the e-folding time of the main stellar population in Myr. When the SED fitting is done without using the information from H$\delta$ and D$_n$4000, for the majority of the galaxies the e-folding time has values $\geq 500$\,Myr. The opposite trend is observed when H$\delta$ and D$_n$4000 are included in the fitting process. In this case, most of the galaxies have short e-folding times ($\leq 200$\,Myr). The bottom panel presents the mass fraction of the late burst population. Inclusion of the two indices does not significantly affect the calculated values of this parameter.}
  \label{fig_sfh}
\end{figure} 
 
\begin{figure}
\centering
  \includegraphics[width=0.75\columnwidth, height=5.8cm]{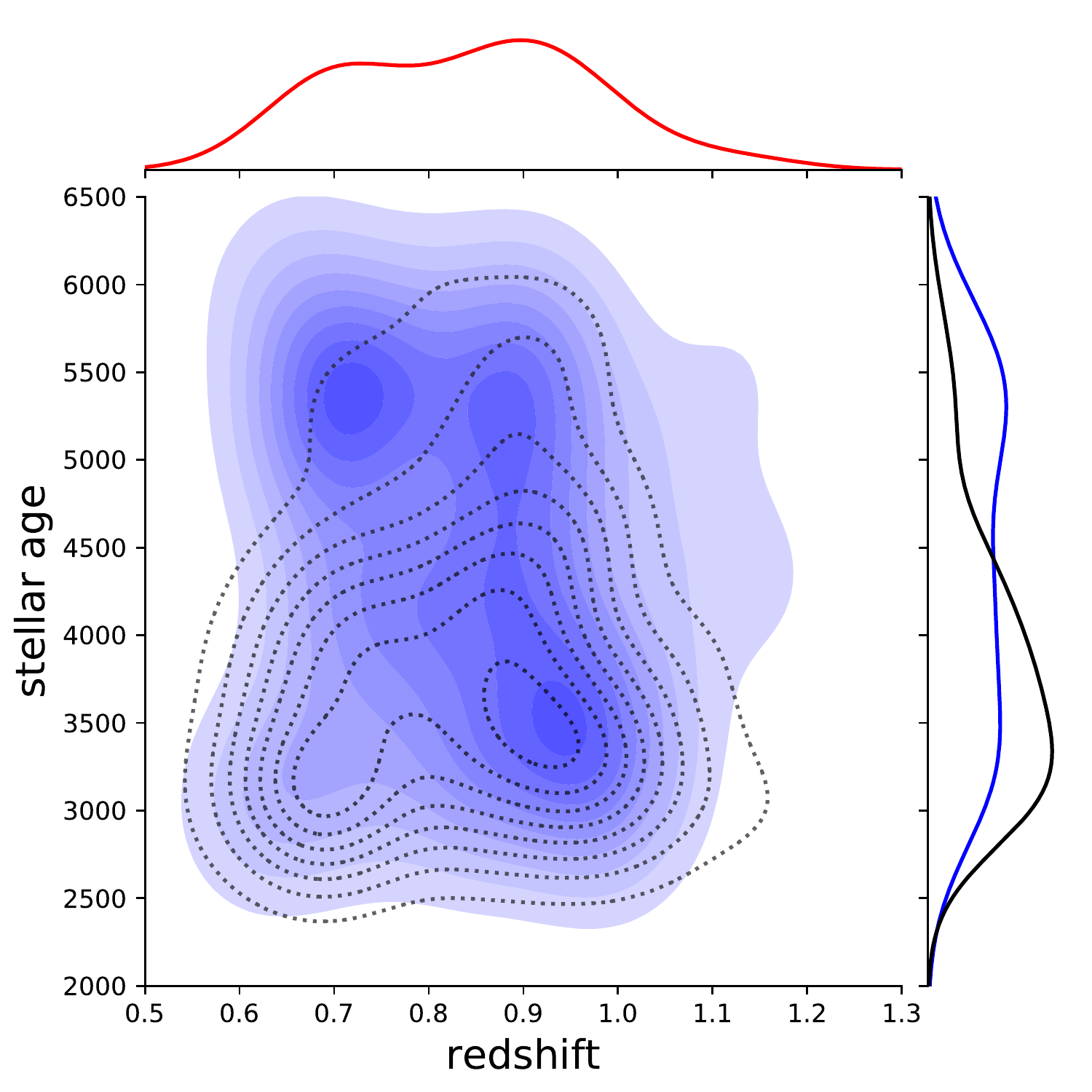}   
  \includegraphics[width=0.75\columnwidth, height=5.8cm]{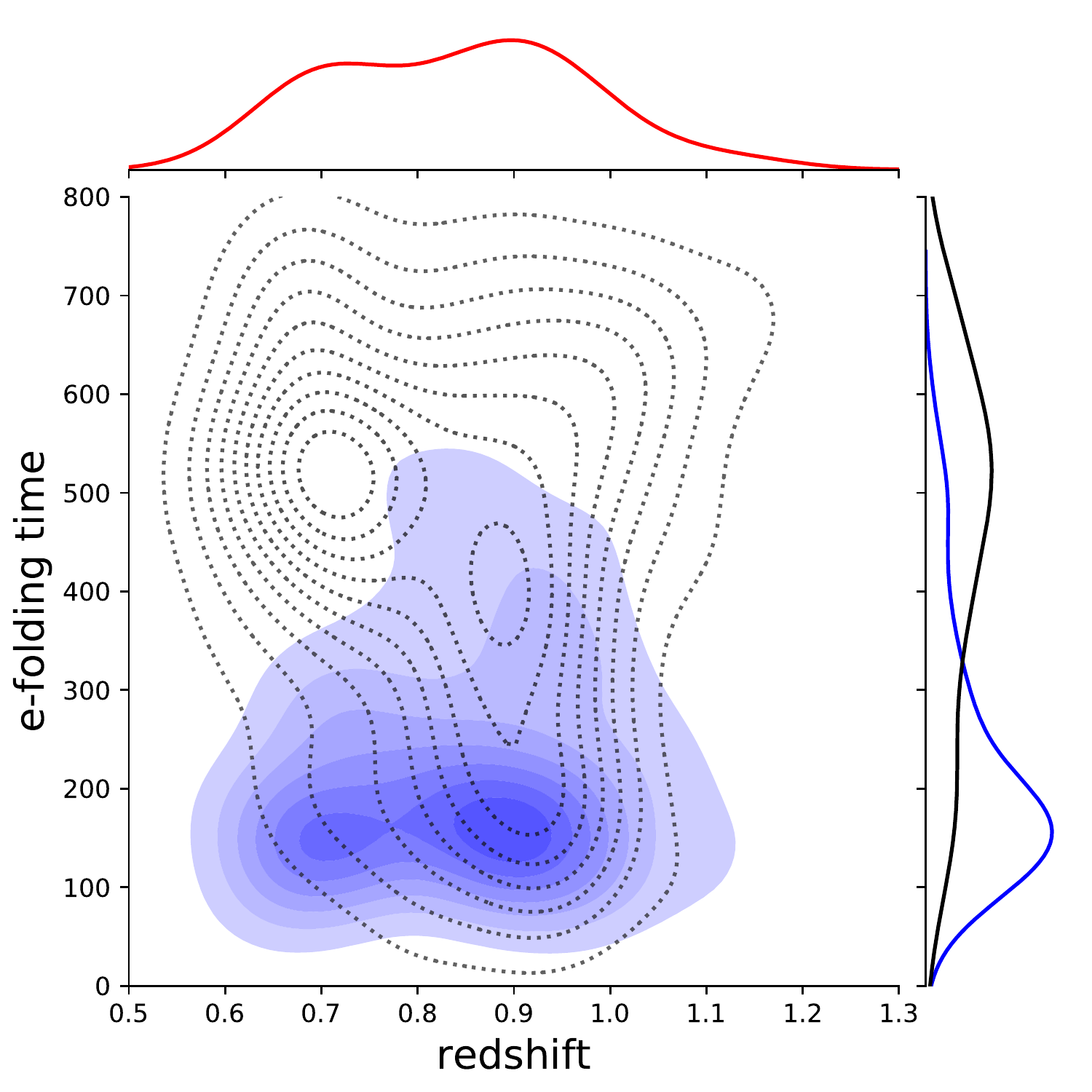} 
  \includegraphics[width=0.75\columnwidth, height=5.8cm]{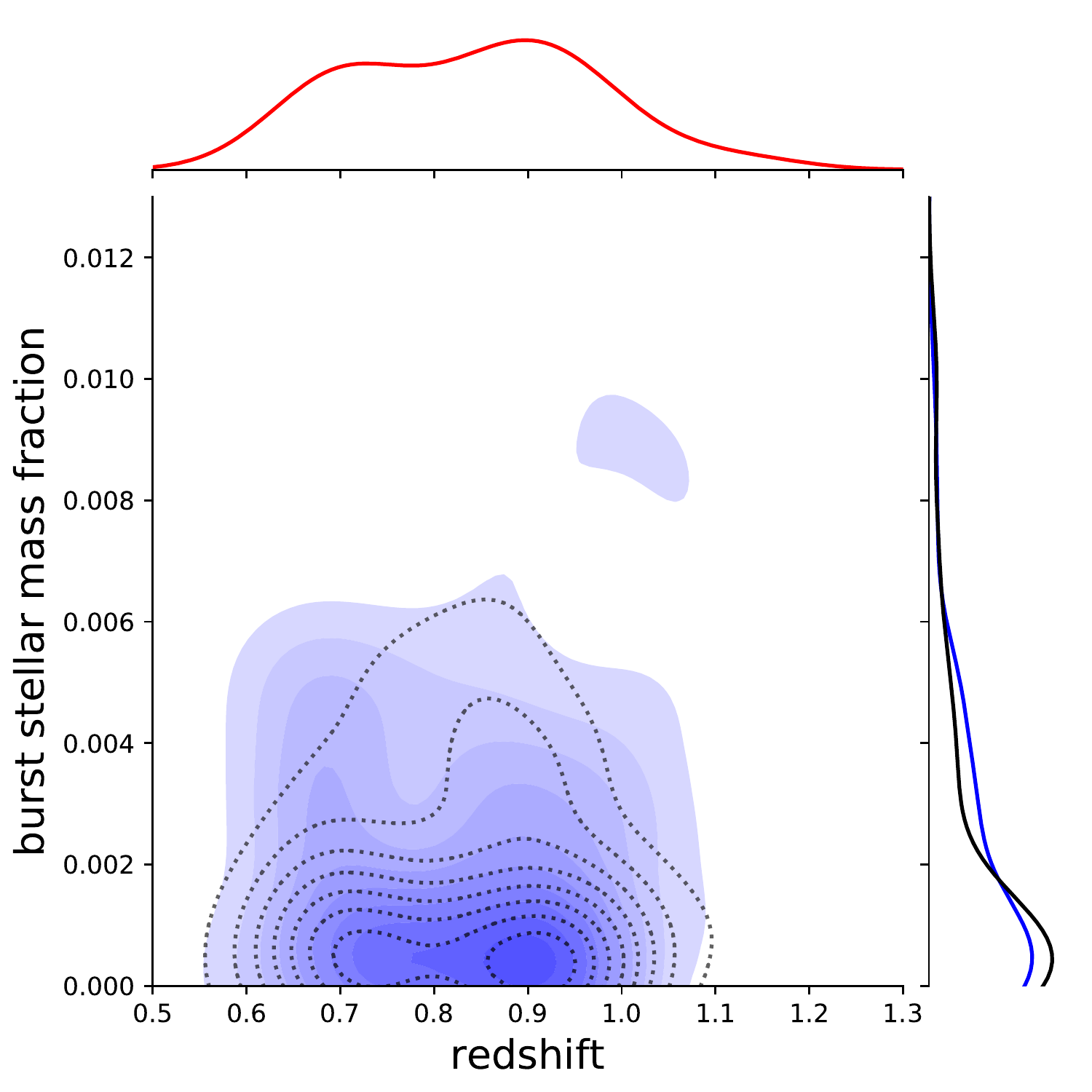}
  \caption{Same format as in Fig. \ref{fig_sfh}, but for the AGN sample. Similar trends are observed with those for the reference catalogue.}
  \label{fig_sfh_agn}
\end{figure}  

\begin{figure*}
\centering
\begin{subfigure}{.50\textwidth}
  \centering
  \includegraphics[width=.9\linewidth]{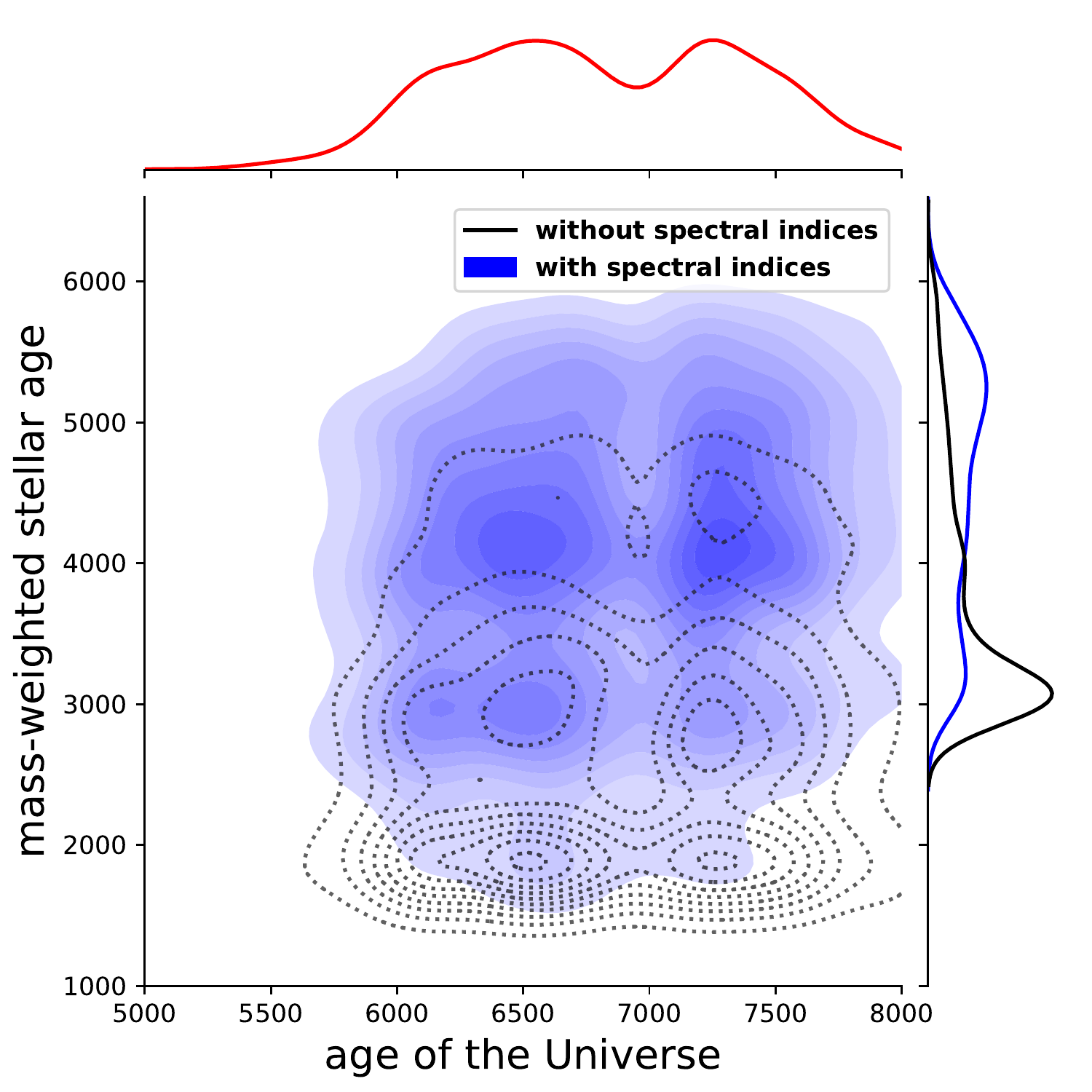}
  \label{z}
\end{subfigure}%
\begin{subfigure}{.50\textwidth}
  \centering
  \includegraphics[width=.9\linewidth]{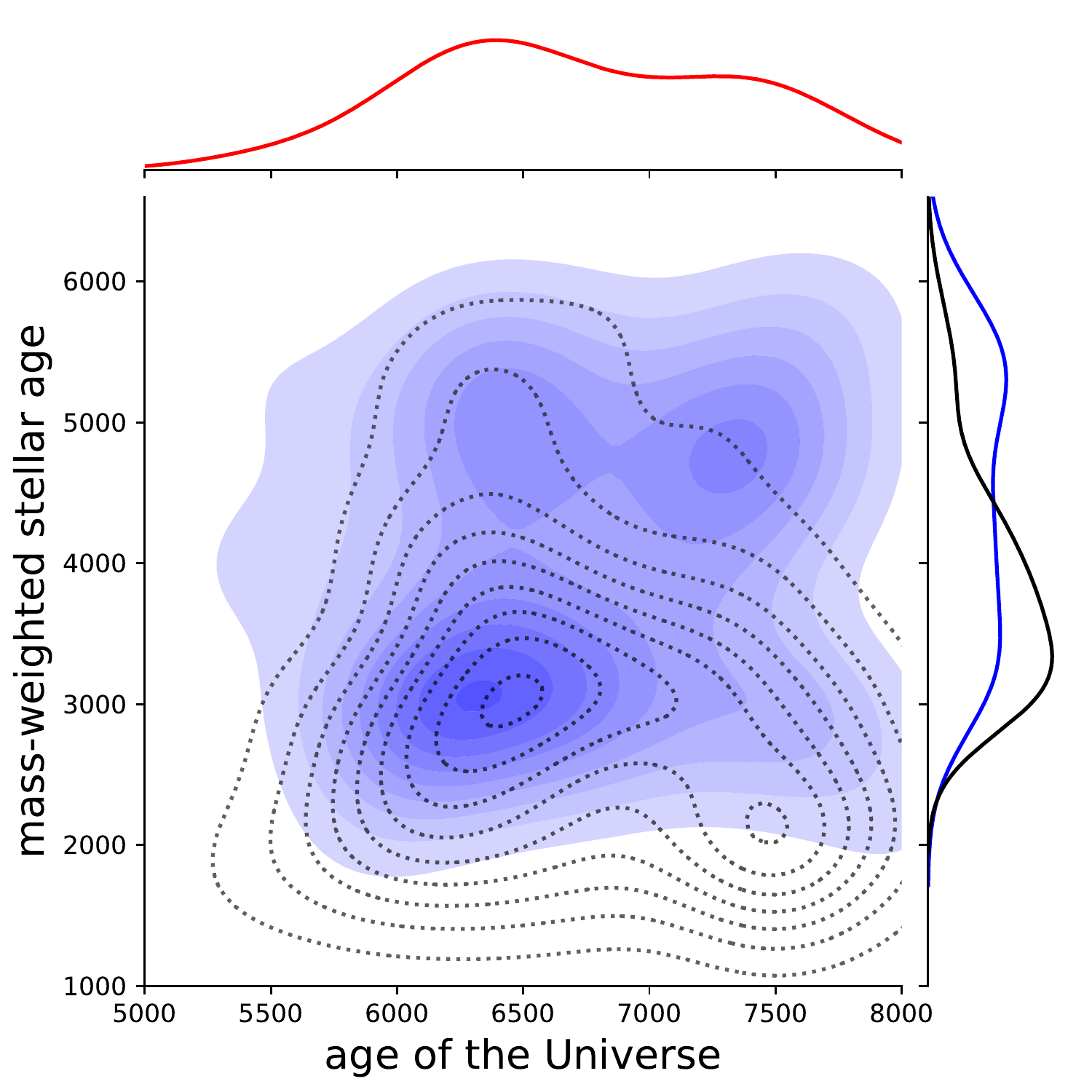}
  \label{Lx}
\end{subfigure}
\caption{Comparison of the mass-weighted stellar ages (in Myr) calculated by CIGALE, with (blue shaded contours) and without (black line contours) using H$\delta$ and D$_n$4000 in the fitting process, for sources in the reference catalogue (left panel) and AGN (right panel). The addition of the two indices in the SED fitting provides additional constraints on the stellar population and allows to CIGALE to calculate, on average, more meaningful stellar ages, in the sense that the distributions of the mass-weighted stellar age are flatter and resemble better the distribution of the age of the Universe, as compared to the highly peaked (at low values) mass-weighted stellar age distributions without the spectral indices.}
\label{fig_agemain_vs_redz}
\end{figure*}

\begin{figure}
\centering
  \includegraphics[width=0.9\columnwidth, height=6.5cm]{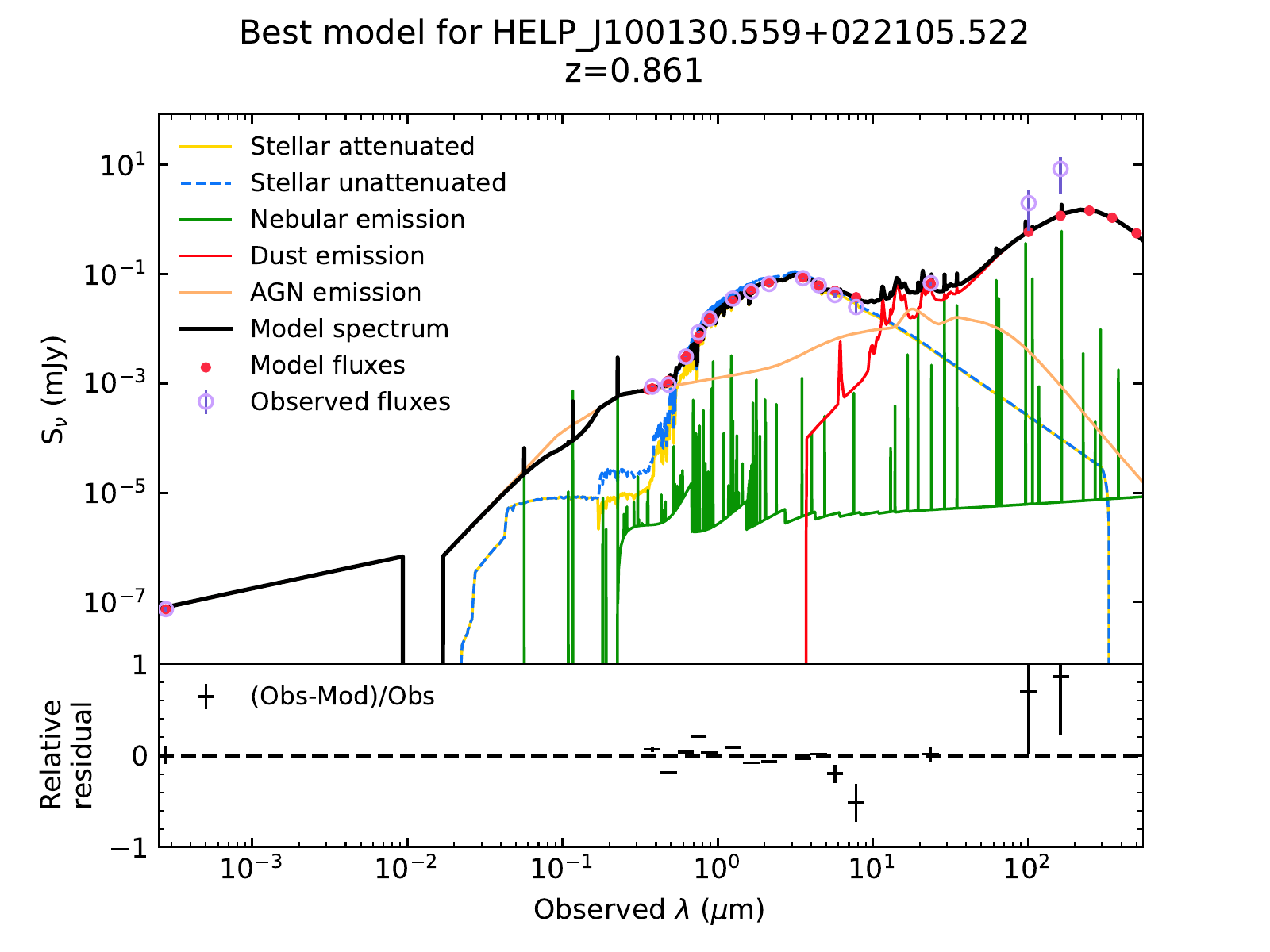} 
  \includegraphics[width=0.9\columnwidth, height=6.5cm]{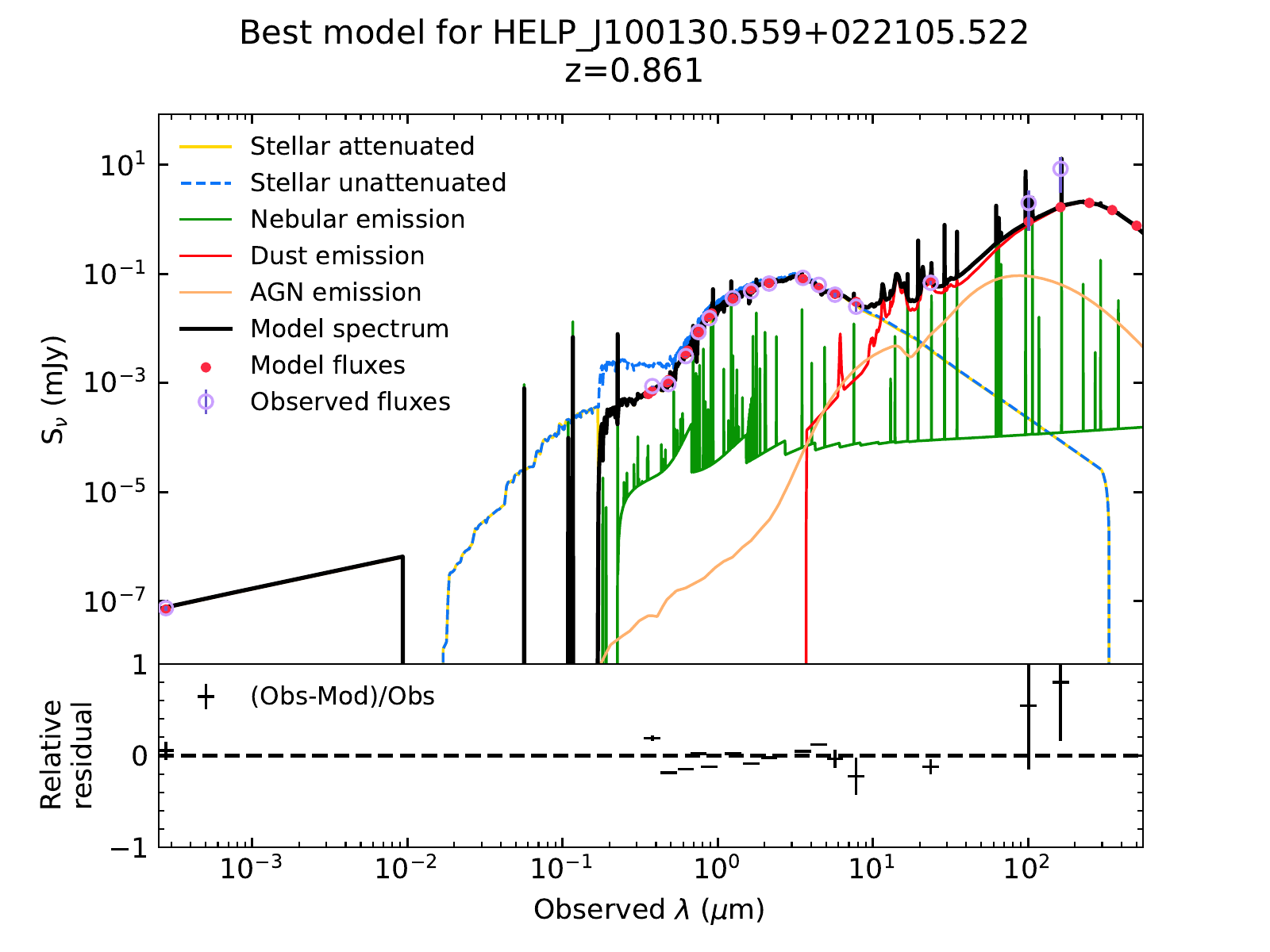} 
  \caption{Example of an AGN SED, when the two spectral indices are included in the fitting process (top panel) and when they are not included (bottom panel).}
  \label{fig_seds}
\end{figure}

\begin{figure}
\centering
  \includegraphics[width=0.9\columnwidth, height=6.5cm]{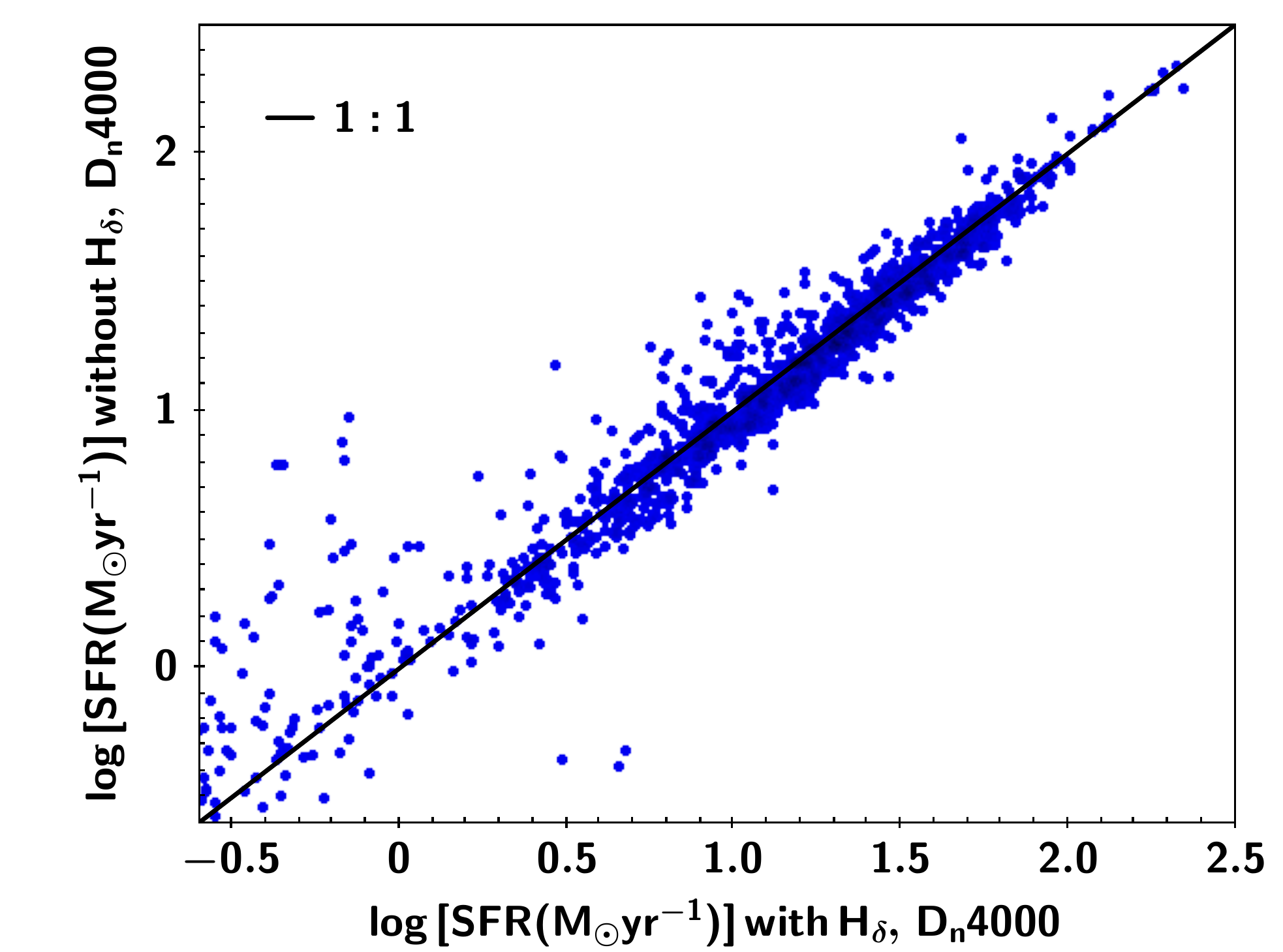} 
  \includegraphics[width=0.9\columnwidth, height=6.5cm]{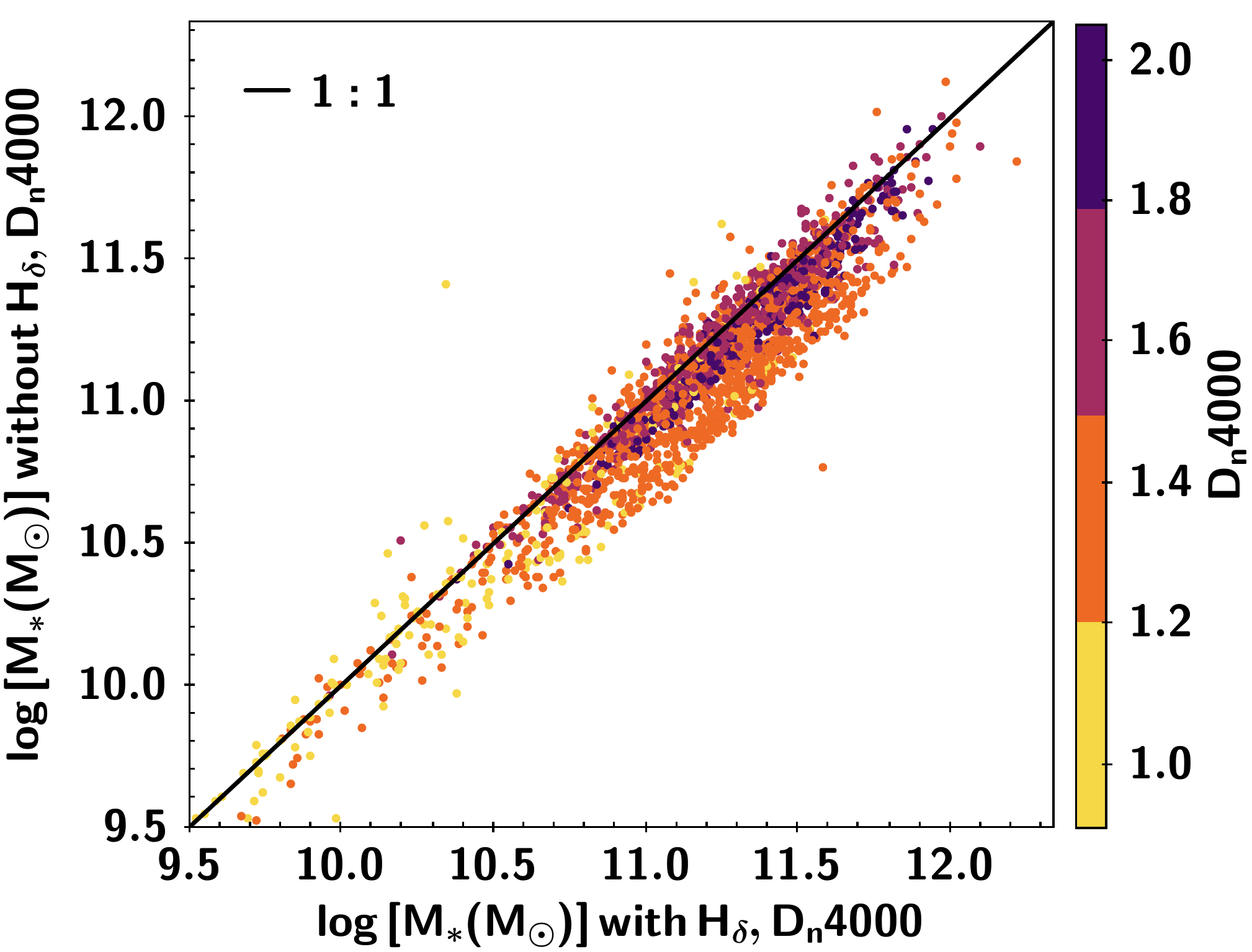} 
  \caption{Comparison of galaxy properties with and without H$\delta$ and D$_n$4000 in the fitting process, for sources in the reference catalogue. The top panel compares the SFR calculations of CIGALE with and without the two indices in the SED fitting. The two measurements are in good agreement, indicating that the inclusion of H$\delta$ and D$_n$4000 does not significantly affect the SFR calculations. The bottom panel presents the comparison of the stellar mass measurements. In this case, inclusion of the two indices in the SED fitting causes a small (by $\sim0.2$\,dex), but systematic increase of M$_*$. This is more evident for systems that host stars with young populations (D$_n$4000$<1.5$). Solid, black lines present the 1:1 relation.}
  \label{fig_gal_properties}
\end{figure} 

\begin{figure}
\centering
  \includegraphics[width=0.9\columnwidth, height=6.5cm]{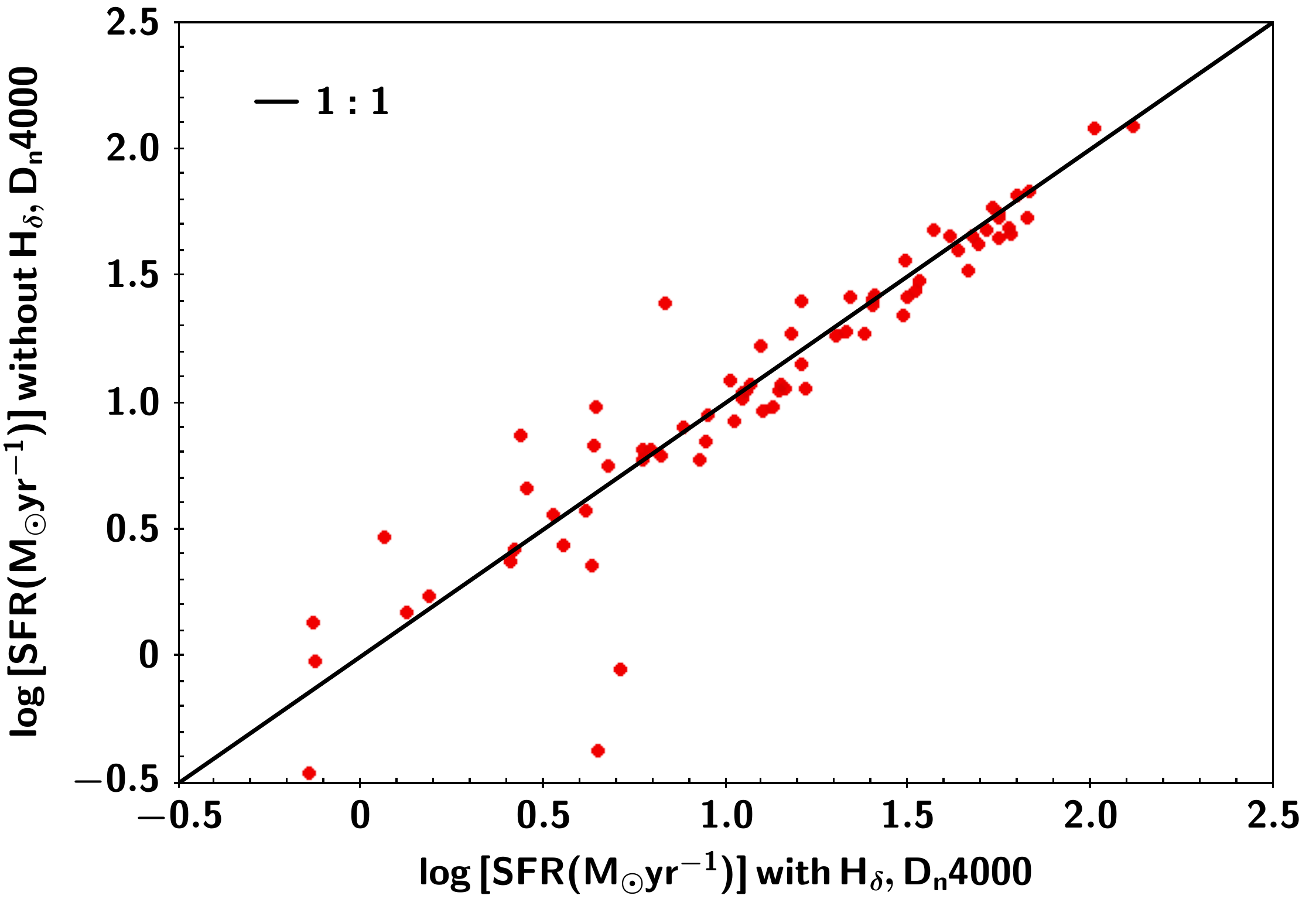} 
  \includegraphics[width=0.9\columnwidth, height=6.5cm]{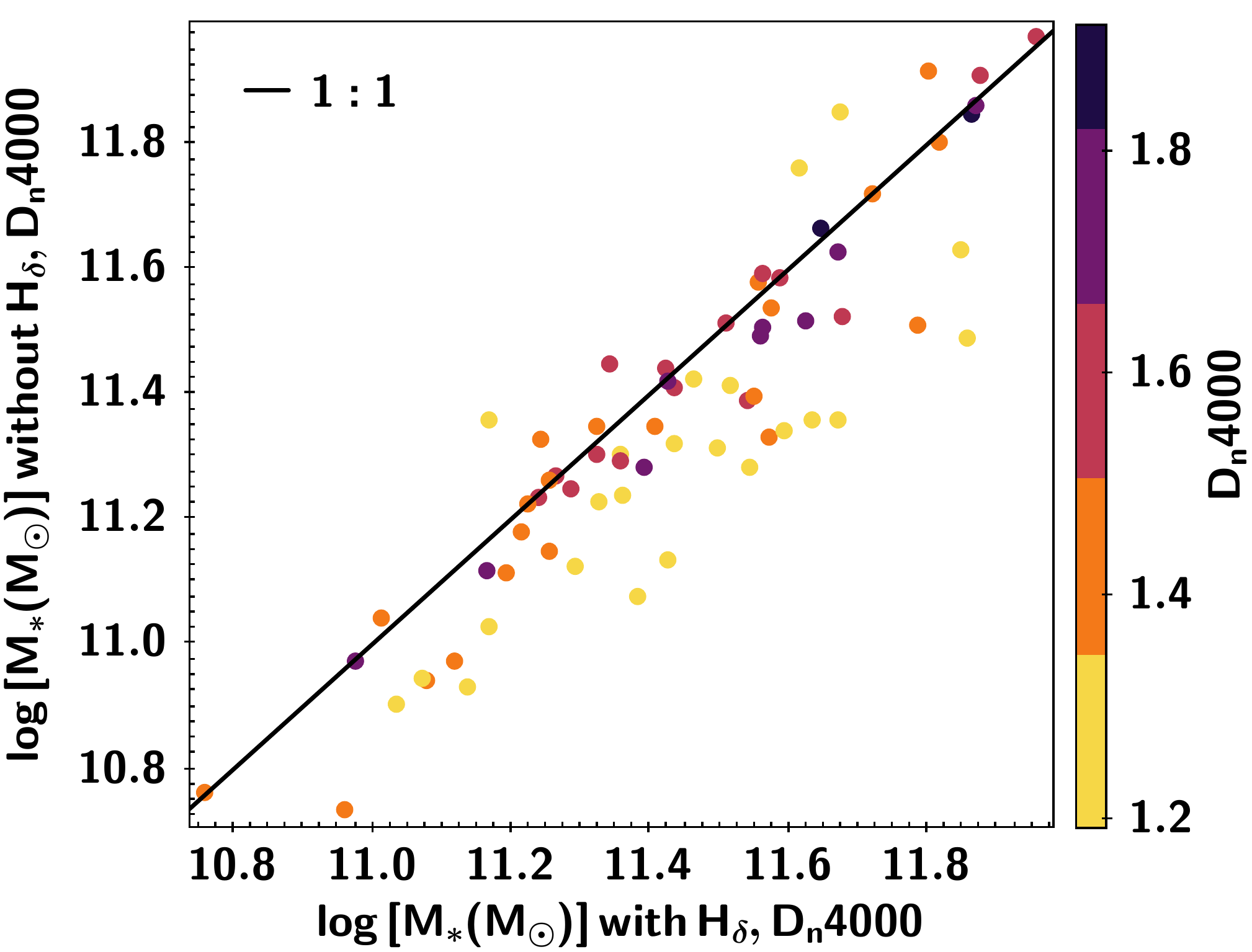} 
  \caption{Same format as in Fig.\ref{fig_gal_properties}, but for galaxies that host AGN.  Similar trends are observed with those in Fig.\ref{fig_gal_properties}.}
  \label{fig_gal_properties_agn}
\end{figure}

\begin{figure}
\centering
  \includegraphics[width=0.9\columnwidth, height=6.5cm]{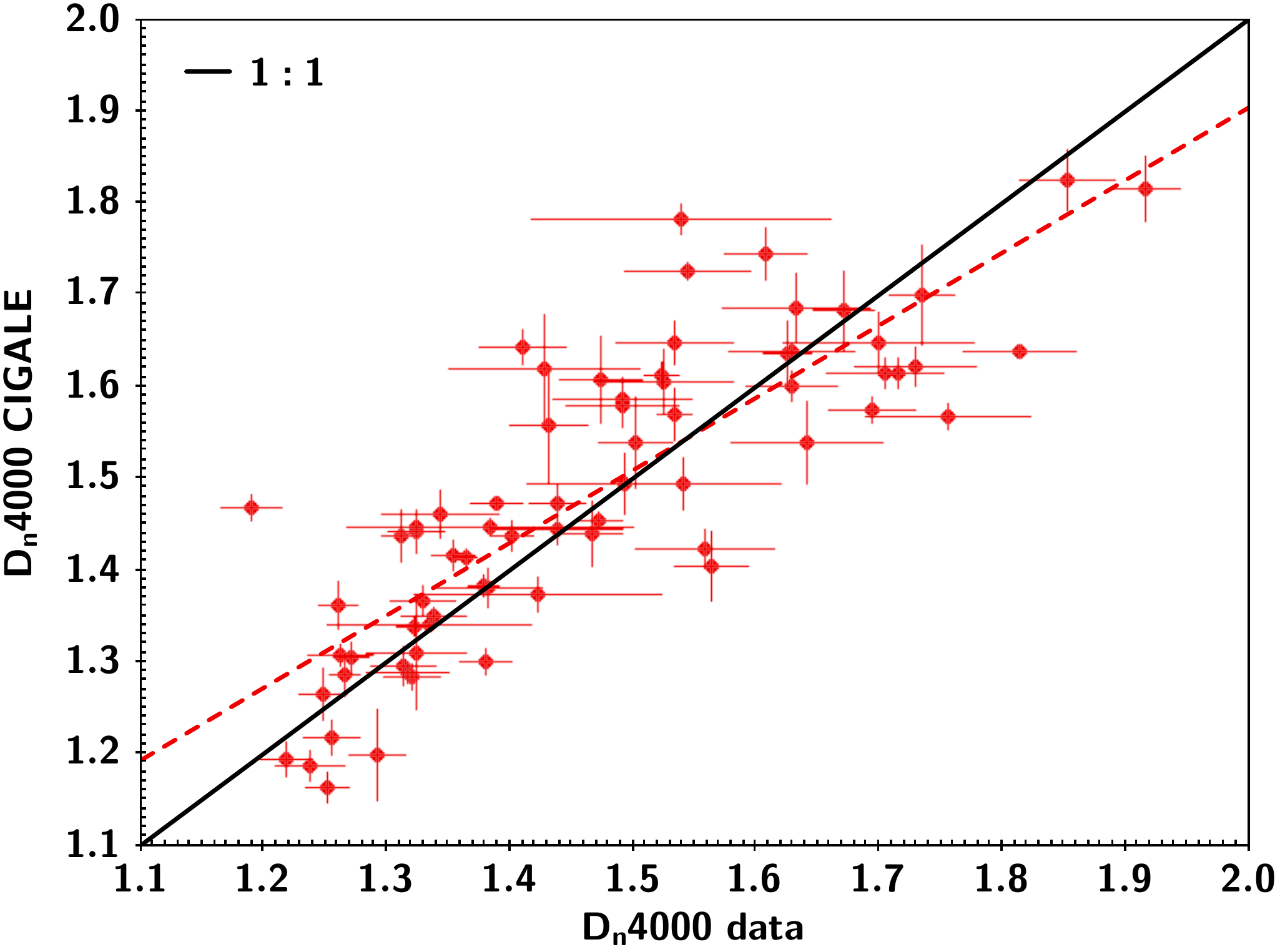} 
  \includegraphics[width=0.9\columnwidth, height=6.5cm]{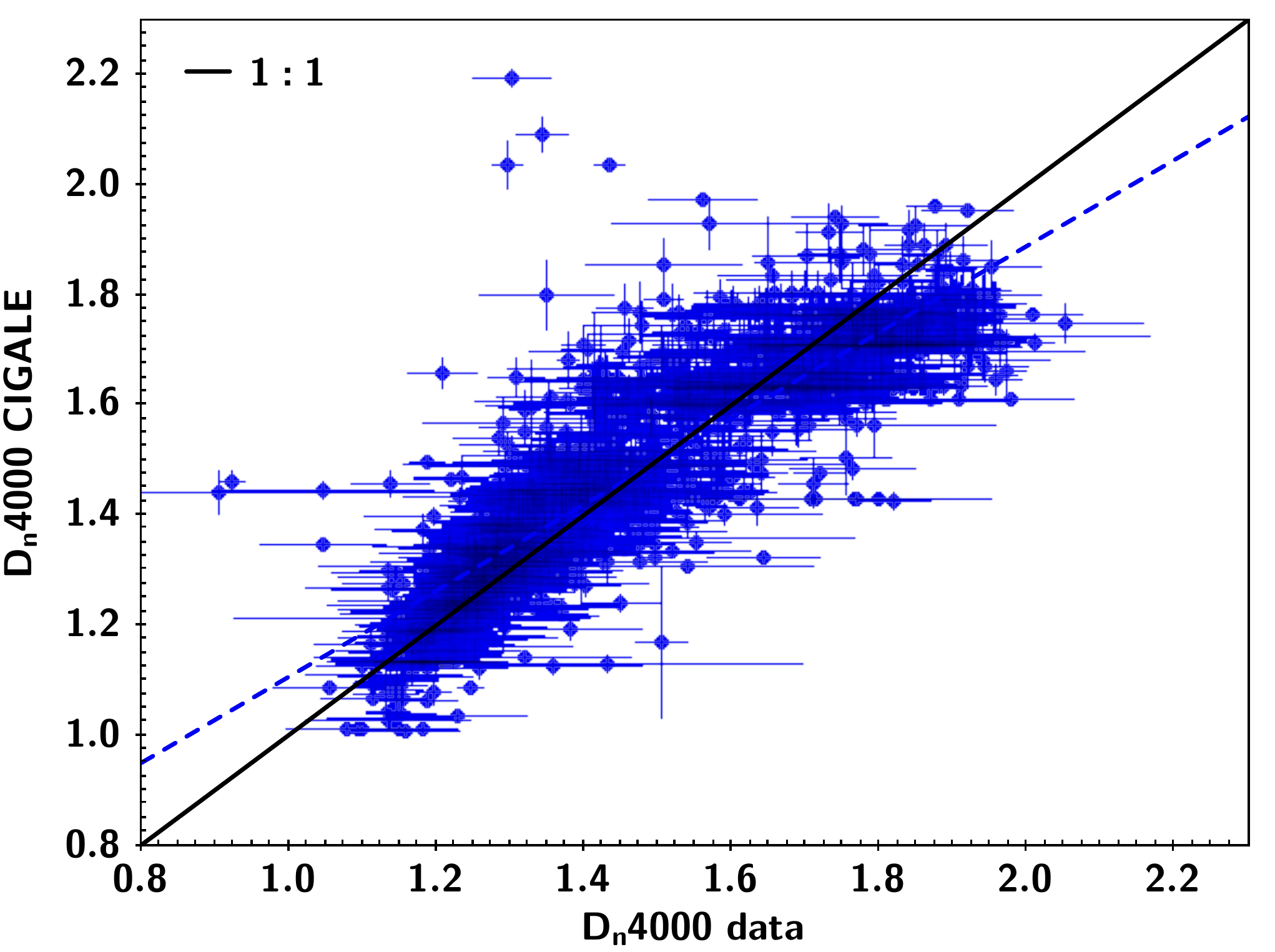} 
  \caption{Comparison of the D$_n$4000 measurements of CIGALE to the input (data) values for the AGN (top panel) and the galaxies in the reference catalogue (bottom panel). The index is not included as input (data) in the fitting process. In both cases, the algorithm successfully recovers the input value of the index. The dashed lines in the two panels, present the best linear fits. In the case of AGN the fit is given by the following expression: D$_n$4000$\rm_{CIGALE}=0.7903\times$ D$_n$4000$\rm_{data}+0.3236$. For the reference sample a similar equation is found: D$_n$4000$\rm_{CIGALE}=0.7567\times$ D$_n$4000$\rm_{data}+0.3230$. Solid, black lines present the 1:1 relation.}
  \label{fig_D4000_calibration}
\end{figure}

\begin{figure}
\centering
  \includegraphics[width=0.9\columnwidth, height=6.5cm]{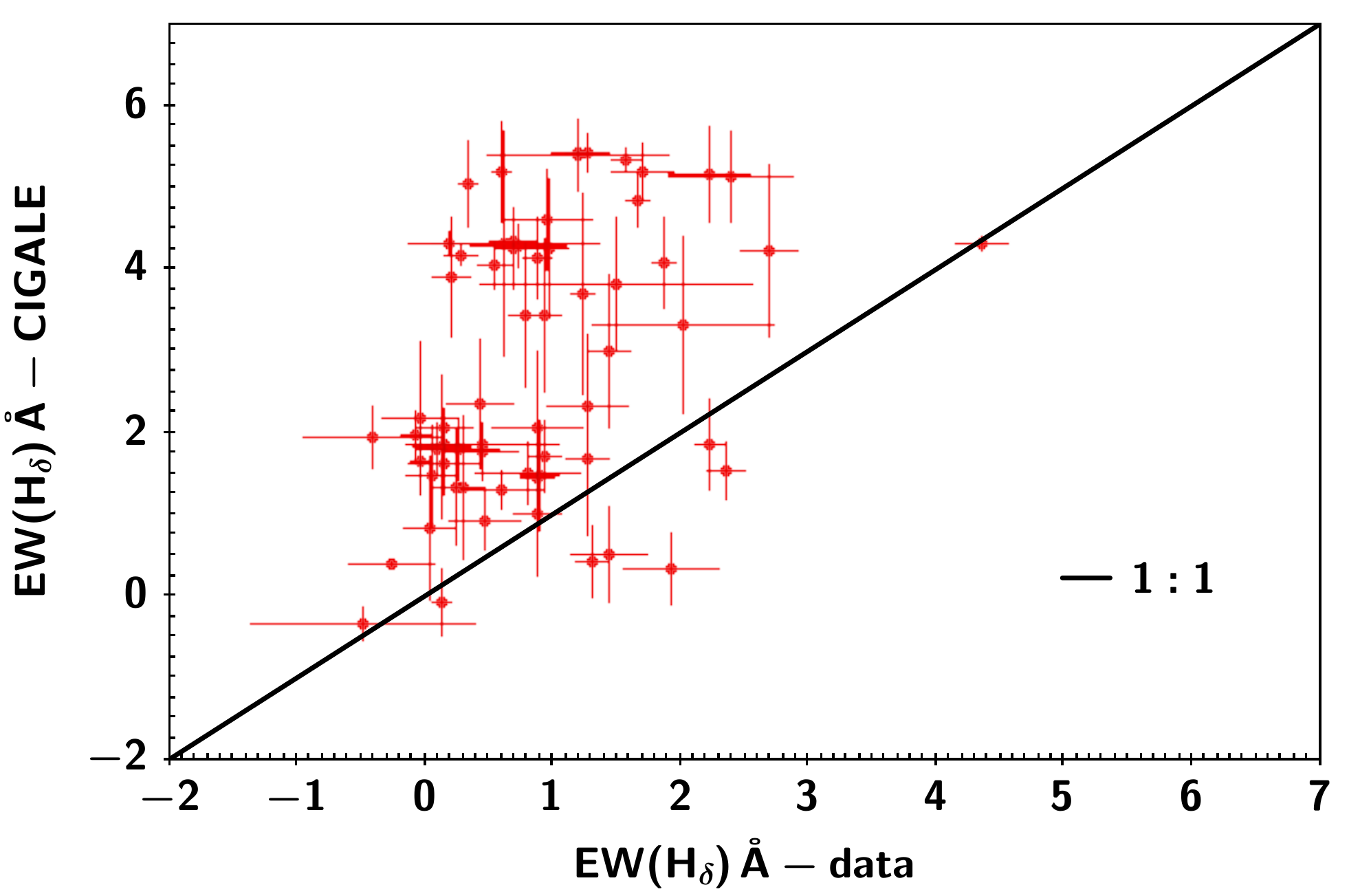} 
  \includegraphics[width=0.9\columnwidth, height=6.5cm]{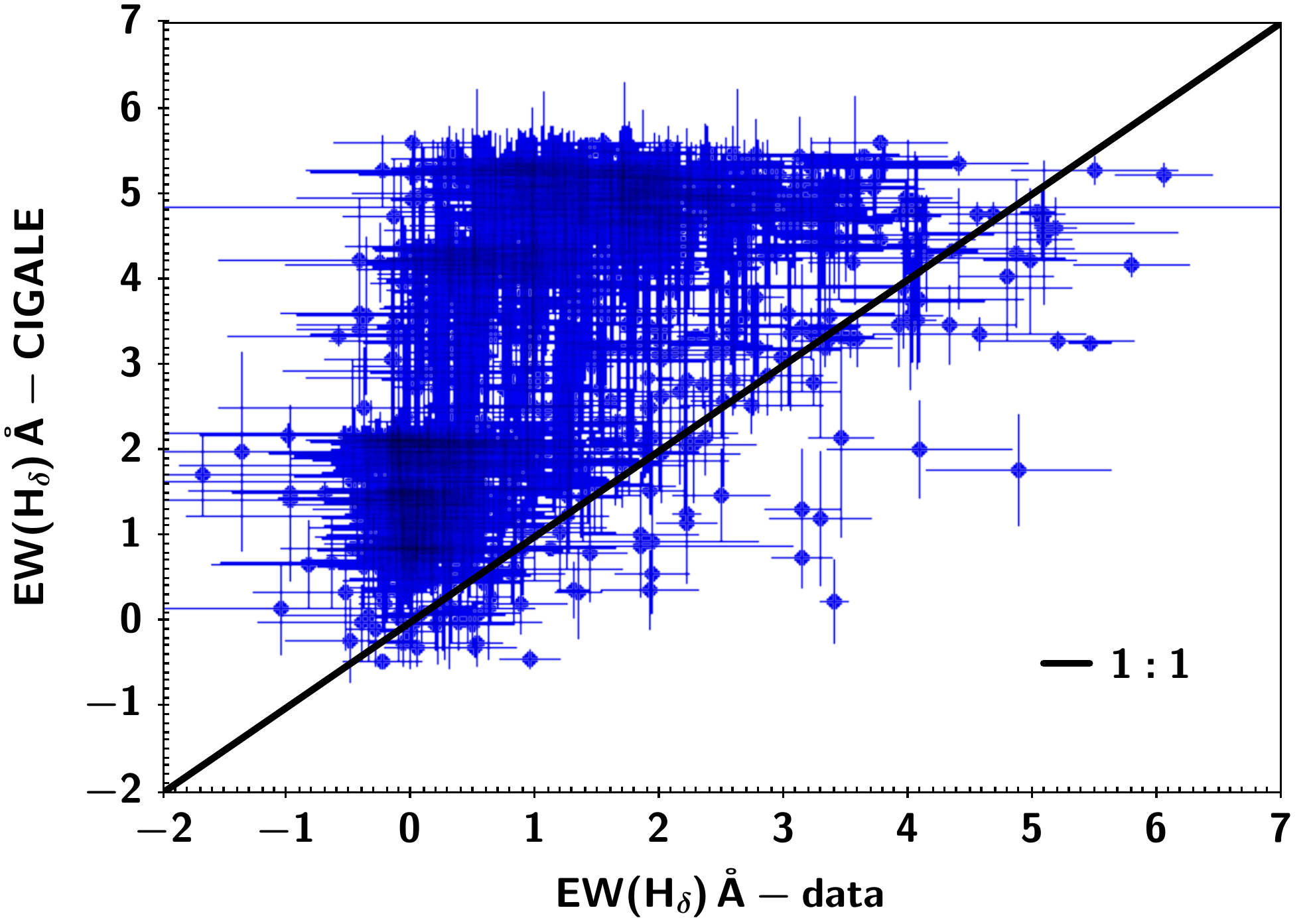} 
  \caption{Comparison of the H$\delta$ measurements of CIGALE to the input (data) values for the AGN (top panel) and the galaxies in the reference catalogue (bottom panel), when the line is not used as an input in the fitting process. The algorithm tends to overestimate the true values, in particular for EW higher than $\sim 0.1$. Solid, black lines present the 1:1 relation.}
  \label{fig_hdelta_calibration}
\end{figure}


From the 94 AGN, 69 ($\approx 75\%$) have available measurement for the D$_n$4000 index and 83 ($\approx 85\%$) have measurement of the EW of the H$\delta$ absorption line. All AGN with available D$_n$4000 also have available measurement for H$\delta$. Therefore, 69/94 AGN have measurements for both indices. The corresponding numbers for the galaxy sample are: $\approx 82\%$ of galaxies have a measurement of H$\delta$ and $\approx 75\%$ a measurement for D$_n$4000. Approximately, $75\%$ (2176) of the sources in the reference catalogue have available measurements for both indices.

\section{The impact of absorption lines on the SED fitting measurements}
\label{sec_cigale}

SFH parameters suffer from large degeneracies and is difficult to accurately constrain with any SED fitting algorithm \citep[e.g.,][]{Ciesla2016, Chisholm2019}. In this section, we examine if the inclusion of the H$\delta$ and D$_n$4000 indices in the SED fitting process affects the reliability with which the algorithm estimates the SFH parameters, if it improves their calculations and whether it affects the measurements of the (host) galaxy properties. 



\begin{figure}
\centering
  \includegraphics[width=0.9\columnwidth, height=6.5cm]{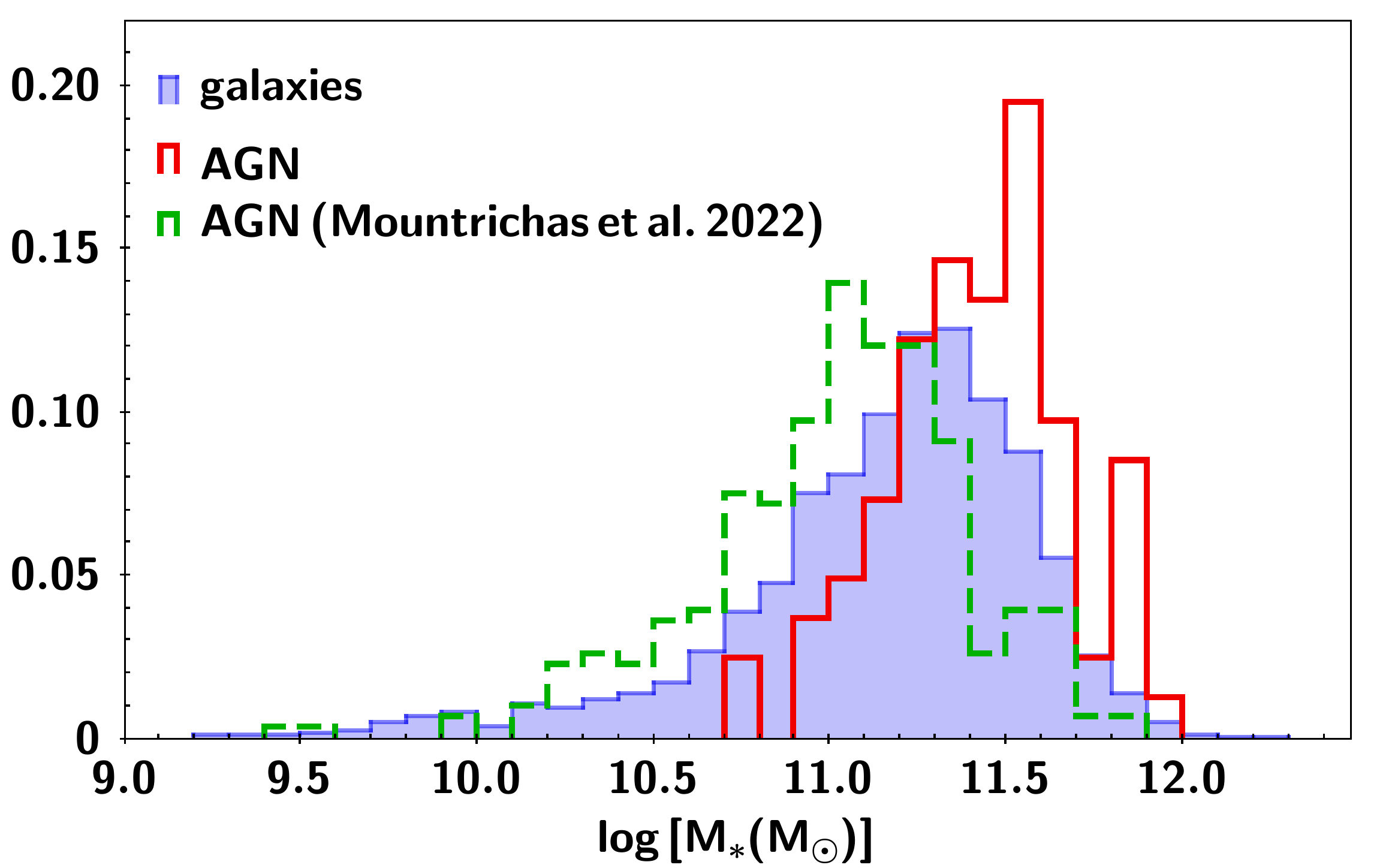} 
  \caption{Stellar mass distributions of sources in the reference galaxy catalogue (blue shaded histogram), AGN used in this study (red line) and the AGN sample used in \cite{Mountrichas2022a} (green line).}
  \label{fig_mstar}
\end{figure}

For this exercise, we use the CIGALE SED fitting algorithm \citep{Boquien2019, Yang2020, Yang2022}. CIGALE allows inclusion of the X-ray flux in the fitting process and has the ability to also account for the extinction of the UV and optical emission in the poles of AGN \citep{Yang2020, Mountrichas2021a, Mountrichas2021b, Buat2021}. CIGALE is able to combine both photometric data and spectral indices in the same fit.

Table \ref{table_cigale} presents the templates and the values for the free parameters used in the fitting process. The galaxy component is modelled using a delayed star formation history (SFH) model with a function form $\rm SFR\propto t \times exp(-t/\tau$). A star formation burst is included \citep{Ciesla2017, Malek2018, Buat2019} as a constant ongoing period of star formation of 50\,Myr. Stellar emission is modelled using the single stellar population templates of \cite{Bruzual_Charlot2003} and is attenuated following the \cite{Charlot_Fall_2000} attenuation law. The emission of the dust heated by stars is modelled based on \cite{Dale2014}, without any AGN contribution. The AGN emission is included using the SKIRTOR models of \cite{Stalevski2012, Stalevski2016}. 

We have extended CIGALE to compute the H$\delta$ absorption line equivalent width. To do so, we define three wavelength windows. We use the windows at the shortest and longest wavelengths to measure the continuum level of the absorption line. The equivalent width is then computed subtracting the mean level from the central window and integrating the spectrum. We fit equivalent widths to the observations similarly to broadband fluxes, passing the rest-frame equivalent widths and the corresponding uncertainties to the input file in units of nm, along with fluxes in units of mJy.


\begin{figure*}
\centering
  \includegraphics[height=12cm]{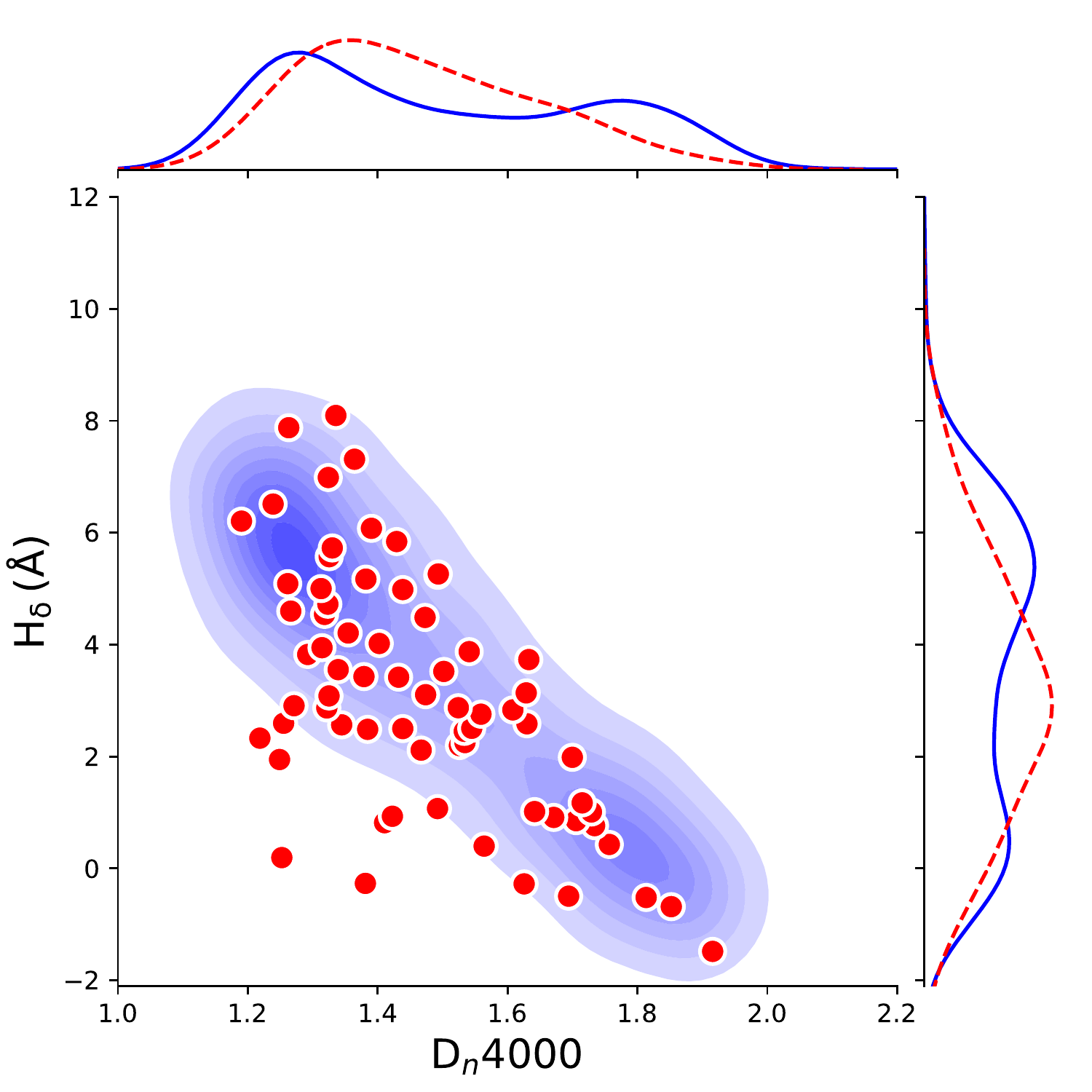} 
  \caption{H$_\delta$ and D$_n$4000 of AGN (green circles) and galaxies (blue contours) in the reference sample. The D$_n$4000 of galaxies appears double peaked with one peak at D$_n$4000$\sim 1.2$ and a second peak at D$_n$4000$\sim 1.8$. On the other hand, AGN present a peak at  D$_n$4000$\sim 1.4$ and a long tail that extends out to D$_n$4000$\sim 1.8$. Regarding H$_\delta$, the distribution of AGN shows a peak at H$\delta \sim3$, while galaxies in the reference catalogue have a flatter distribution.}
  \label{fig_hd_vs_d4000}
\end{figure*}

\begin{figure}
\centering
  \includegraphics[width=0.99\columnwidth, height=8.5cm]{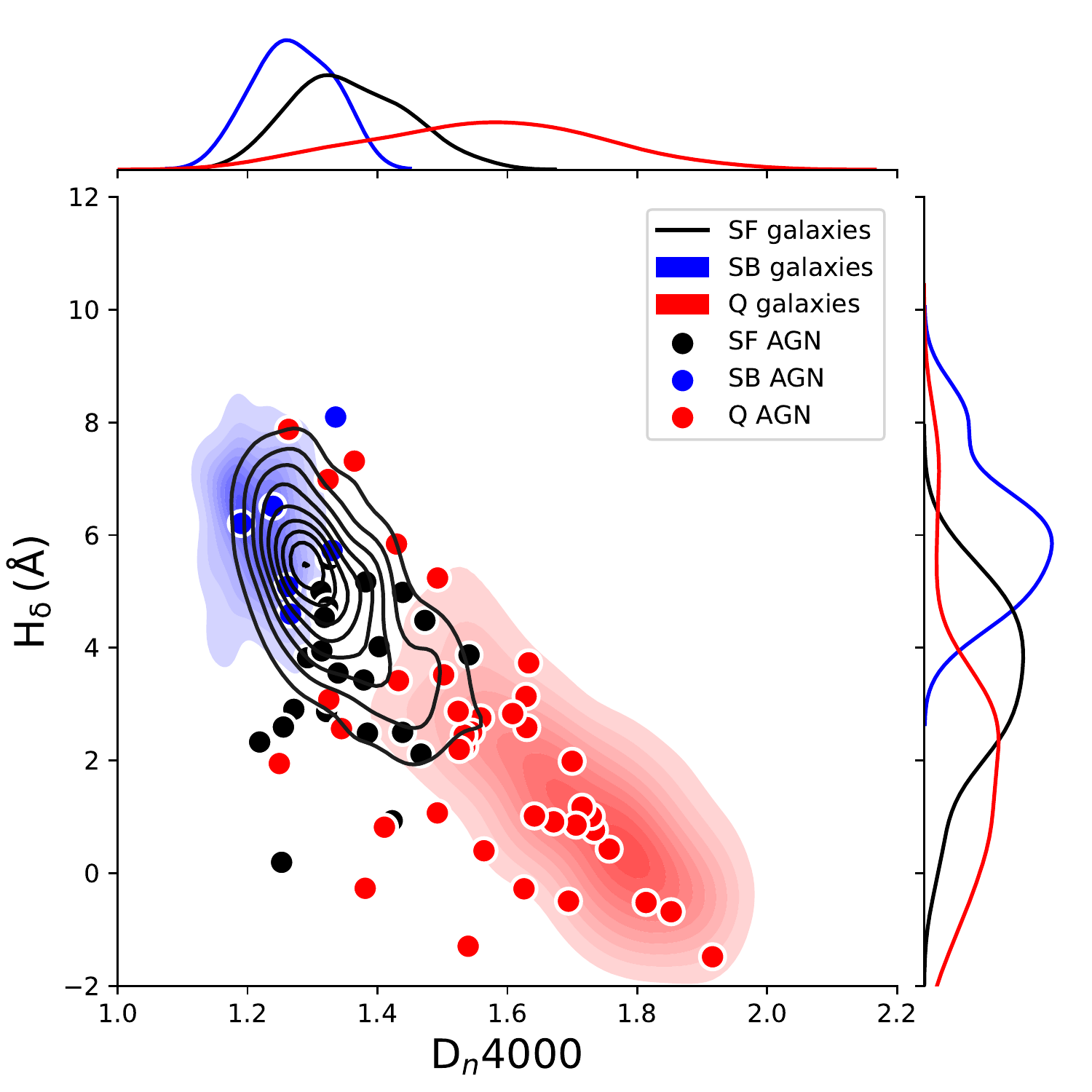} 
  \caption{The location of SB, SF and Q AGN (circles) and galaxies in the reference catalogue (contours) in the H$_\delta$-D$_n$4000 space. The distribution of H$_\delta$ and D$_n$4000 for the AGN population is also presented. }
  \label{fig_hd_vs_d4000_ms}
\end{figure}

\begin{figure}
\centering
  \includegraphics[width=0.99\columnwidth, height=8.5cm]{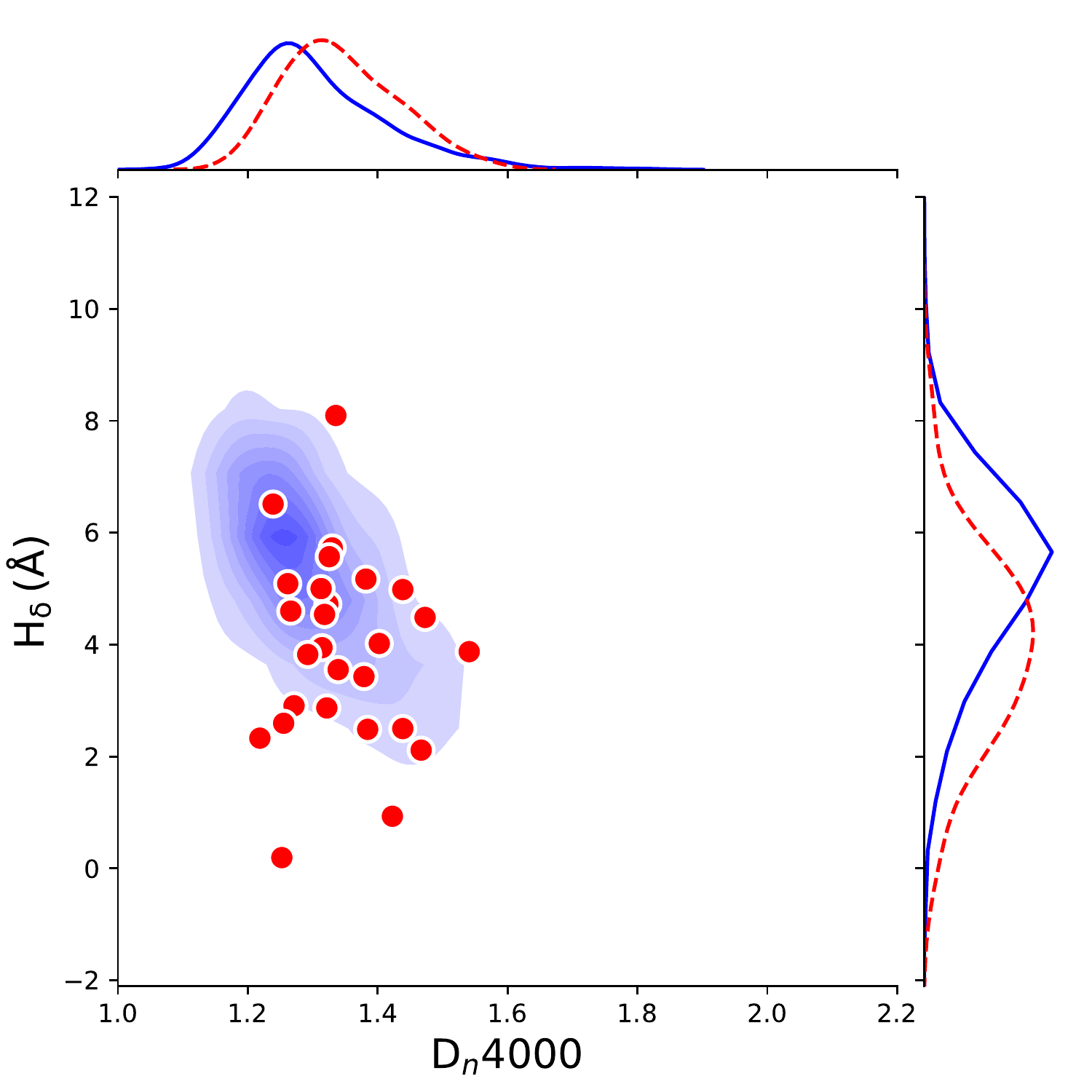} 
  \caption{Same format as in Fig. \ref{fig_hd_vs_d4000}, but excluding quiescent systems using the sSFR distributions of the sources (see text for more details). For both populations the vast majority of sources with high D$_n$4000 values (D$_n$4000$>1.6$) has been excluded. In the case of the  H$_\delta$ distributions, the exclusion of quiescent systems results in the exclusion of the majority of sources with  H$_\delta<2$. However, there is still a significant fraction of AGN with H$_\delta \sim 2-4$. On the other side of the distribution, $\sim 20\%$ of the galaxies have H$_\delta>6$, whereas the corresponding fraction of AGN is $\sim 5\%$.} 
  \label{fig_hd_vs_d4000_excl_quiescent}
\end{figure}

\subsection{Effect on galaxy properties}

In this section, we examine the effect that the inclusion of the two indices in the SED fitting process has on the calculation of the SFH parameters and on the (host) galaxy properties.


First, we examine whether the parameter space used in the SED fitting, described in the previous section (see also Table \ref{table_cigale}), sufficiently covers the data space of the H$_\delta$ and D$_n$4000 indices. For this purpose, we use the models built by CIGALE prior to the fitting process. CIGALE creates models using the templates and parameter space and determines the full list of parameters for each model to be computed. Then, the spectrum is computed for each model, as well as its physical properties and passbands \citep[for more details see Sect. 4.2 in][]{Boquien2019}. In Fig. \ref{models_vs_data}, we plot the EW of H$\delta$ against D$_n$4000 for the models and the AGN and sources in the reference catalogue. The data present lower H$\delta$ values compared to those produced by the models, however, taking into account the statistical uncertainties of the data values, we conclude that the parameter space we use for the SED fitting covers well the data space. 

Figures \ref{fig_sfh} and \ref{fig_sfh_agn} present the distributions of the three, free SFH parameters (Table \ref{table_cigale}), with and without including the H$\delta$ absorption line and the D$_n$4000 index in the SED fitting, for the sources in the galaxy reference catalogue and AGN, respectively. The top panels in the two figures present the distributions of the stellar age in Myr. When the two indices are not included in the fitting process, the distributions highly peak at the lowest stellar age value, allowed by the parametric grid (see Table \ref{table_cigale}). This is at odds with the redshift distribution of the galaxies, presented in Fig. \ref{fig_redz}. Based on the redshift distributions, most of the sources in our sample lie at low(er) redshift ($0.6-0.8$) and thus we expect to have old(er) stellar populations. When the two indices are included in the fitting process, the distribution of stellar ages appears flatter and in better agreement with the redshift distribution. Fig. \ref{fig_agemain_vs_redz}, compares the mass-weighted stellar ages calculated by CIGALE, with and without using H$\delta$ and D$_n$4000 in the fitting process. Again, the inclusion of the two indices in the SED fitting provides additional constraints on the stellar population and allows CIGALE to calculate, on average, more meaningful stellar ages, as the mass-weighted stellar age distributions are flatter and similar to the distribution of the age of the Universe. 

\begin{figure}
\centering
\includegraphics[width=0.99\columnwidth, height=8.5cm]{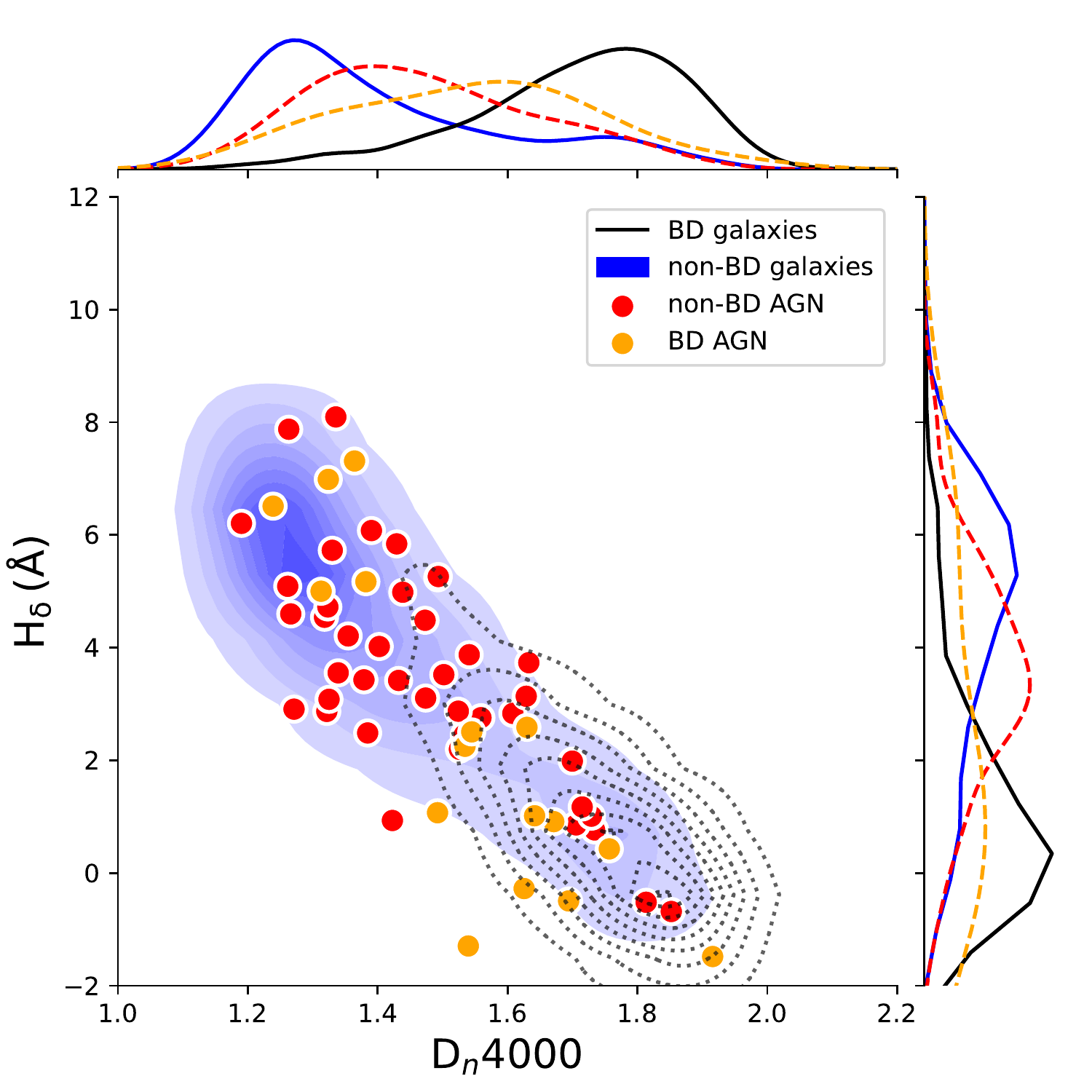} 
  \caption{D$_n$4000 and H$_\delta$ (in \AA) distributions of AGN and galaxies in the reference catalogue, for different morphologies. Non-BD AGN tend to have higher D$_n$4000 values compared to non-BD galaxies that do not host an AGN. However, the opposite trend is observed for BD systems, in which case AGN tend to have lower D$_n$4000 values compared to their non-AGN counterparts. Regarding the H$\delta$, non-BD AGN tend to have lower H$_\delta$ values compared to non-BD, non-AGN galaxies, while BD AGN have a flat H$_\delta$ distribution whereas BD AGN tend to have high H$\delta$ values.} 
  \label{d4000_hdelta_bd}
\end{figure}

The middle panels of Figures \ref{fig_sfh} and \ref{fig_sfh_agn} present the distributions of the e-folding time of the main stellar population in Myr. When the SED fitting is done without using the information from H$\delta$ and D$_n$4000, for the majority of the (host) galaxies the e-folding time has values $\geq 500$\,Myr. The opposite trend is observed when H$\delta$ and D$_n$4000 are included in the fitting process. In this case, most of the galaxies have short e-folding times ($\leq 200$\,Myr). 

The bottom panels of Figures \ref{fig_sfh} and \ref{fig_sfh_agn} present the mass fraction of the late burst population. In this case, inclusion of the two indices does not significantly affect the calculated values of this parameter. This could be due to the (fixed) value of the age of the burst (see Table \ref{table_cigale}) which is too short (50\,Myrs) to be detected by the H$_\delta$ line. Therefore, the inclusion of this line in the fitting process does not provide an additional constrain for the calculation of the mass fraction of the late stellar burst. If we leave the age of the burst a free parameter, mock analysis shows that the parameter cannot be reliably constrained by the algorithm.




Fig. \ref{fig_seds} shows an example of an SED for an AGN, when the spectral indices are included in the fitting process (top panel) and when they are not included (bottom panel). This figure illustrates the effect that the inclusion of the H$\delta$ and D$_n$4000 measurements has on the various emission components. When the two indices are not included in the fit, CIGALE measurements for D$_n$4000 and the EW of H$_\delta$ are 1.58 and 0.16\AA, respectively. The corresponding values from the LEGA-C catalogue are 1.63 and 0.03\AA. When we fit the sources including the two indices, CIGALE's calculations for D$_n$4000 and H$_\delta$ are 1.69 and 0.05\AA, respectively. The mass-weighted stellar age is 4512\,Myr and 3087\,Myr, for the runs with and without the indices, respectively. The redshift of the source is 0.86 (age of the Universe is $\sim 6500$\,Myr).

Next, we examine the effect that the inclusion of H$\delta$ and D$_n$4000 indices in the SED fitting may have on the calculation of the galaxy properties and specifically on the stellar mass, M$_*$, and star formation rate, SFR. 

The top panel of Fig. \ref{fig_gal_properties} presents a comparison of the SFR measurements of CIGALE with and without the two indices in the SED fitting, for sources in the reference catalogue. Based on our results, the inclusion of H$\delta$ and D$_n$4000 does not significantly affect the SFR calculations. This is expected, since the burst stellar mass fraction values are not affected by the inclusion of the two indices (bottom panel of Fig. \ref{fig_sfh}).  However, in the case of stellar mass measurements (bottom panel of Fig. \ref{fig_gal_properties}), we observe a systematic increase (by $\sim0.2$\,dex) of the M$_*$ values when H$\delta$ and D$_n$4000 are included in the fitting process. This is more evident for systems that host young stellar populations (D$_n$4000$<1.5$). Similar trends are observed for the AGN population (Fig. \ref{fig_gal_properties_agn}).

In the LEGA-C catalogue there are available measurements for more lines. In a future paper, we will explore the effect of adding more spectroscopic information alongside the photometric data and in relation to the photometric coverage. 

\begin{figure}
\centering
  \includegraphics[width=0.9\columnwidth, height=6.5cm]{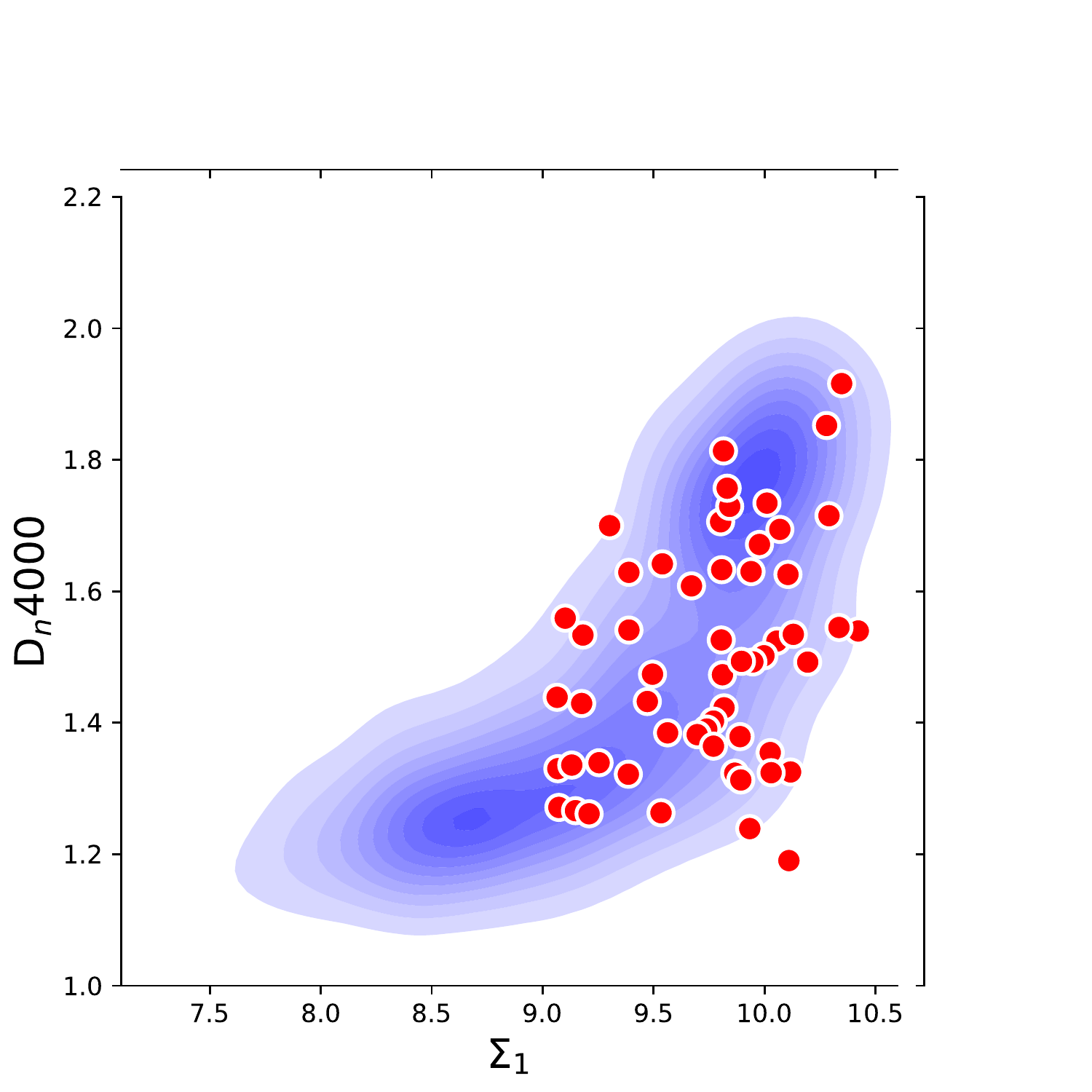} 
  \includegraphics[width=0.9\columnwidth, height=6.5cm]{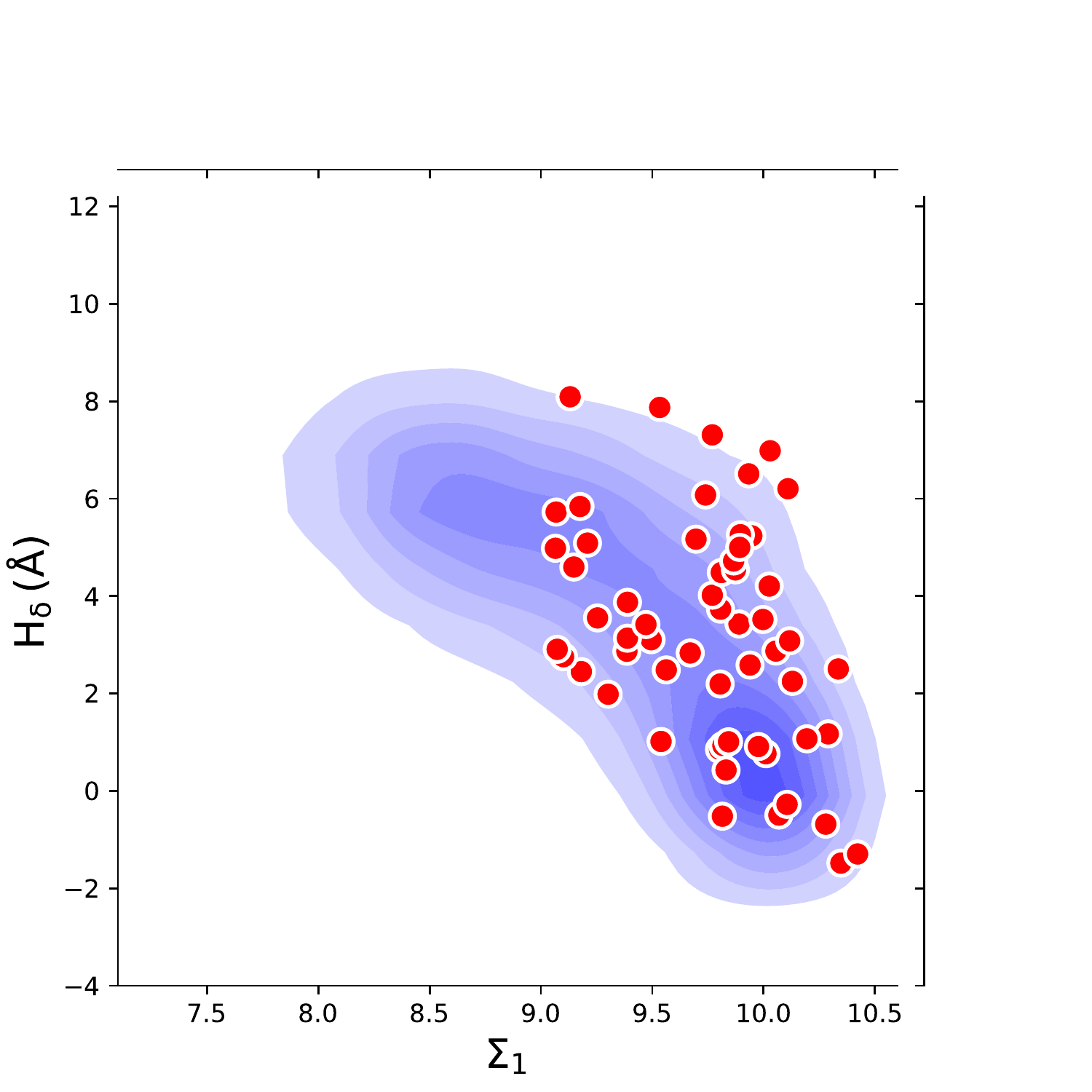} 
  \caption{D$_n$4000 (top panel) and H$_\delta$ (bottom panel) as a function of $\Sigma _1$ (mass-to-size ratio), for AGN (red circles) and galaxies (blue contours). AGN do not have $\Sigma _1$ values lower than 9.0, whereas sources in the reference catalogue go down to 7.5. This implies that AGN reside in more compact systems compared to non-AGN. Both populations present an increase of D$_n$4000 and a decrease of H$_\delta$ with $\Sigma _1$, at $\Sigma _1 \geqslant 9.5$.} 
  \label{d4000_hdelta_S1_both}
\end{figure}

\subsection{Reliability of H$\delta$ and D$_n$4000 calculations from CIGALE}
\label{sec_calibrated}

In this section, we examine how accurately CIGALE can recover the H$\delta$ and D$_n$4000 indices. For that purpose, we use the 69 AGN and the 2,176 sources from the galaxy reference catalogue that have available measurements for both indices in the LEGA-C catalogue and run CIGALE without including the two indices in the fitting process. We then compare CIGALE's calculations for H$\delta$ and D$_n$4000 with those from the LEGA-C catalogue. 

Fig. \ref{fig_D4000_calibration}, compares the D$_n$4000 measurements of CIGALE to the data values, for the AGN (top panel) and the galaxies in the reference catalogue (bottom panel). The index is not included as input in the fitting process. For both populations, the algorithm successfully recovers the value of D$_n$4000. As discussed earlier, this could be due to the extended wavelength coverage of the dataset. The dashed lines in the two panels, present the best linear fits. In the case of AGN the fit is given by the following expression: D$_n$4000$\rm_{CIGALE}=0.7903\times$ D$_n$4000$\rm_{data}+0.3236$. For the reference sample a similar best linear fit is found: D$_n$4000$\rm_{CIGALE}=0.7567\times$ D$_n$4000$\rm_{data}+0.3230$. 


In Fig. \ref{fig_hdelta_calibration}, we present the comparison of CIGALE's measurements with the data values for H$\delta$. In this case, we notice that, both for AGN and galaxies, the algorithm tends to overestimate the true values, in particular for EW higher than $\sim 0.1$\,nm.  This shows that CIGALE cannot predict H$\delta$ using the available photometric coverage. Thus, the inclusion of the line brings additional information in the fitting process and helps the algorithm to break the degeneracies among the SFH parameters to converge to a solution that would not be necessarily selected otherwise. 



In the analysis presented in the next sections, we use the M$_*$ of the AGN and sources in the reference catalogue estimated by the CIGALE runs that include  the two spectral indices in the SED fitting process. 


\section{Comparison of H$\delta$ and D$_n$4000 between AGN and non-AGN systems}

In this section, we compare the values of the H$\delta$ and D$_n$4000 indices between the AGN and sources in the reference catalogue. First, we use the full datasets for both populations, as described in sections \ref{sec_data} and \ref{sec_calibrated}. Then, we restrict the comparison to star forming systems, by excluding quiescent (Q) sources. Finally, we split AGN and galaxies based on their morphology and compactness.

\subsection{Comparison between the full AGN and reference catalogues}

Previous studies that compared the SFR of X-ray AGN with that of star-forming galaxies found that low to intermediate X-ray luminosity AGN, as those used in this study, have SFR that is lower or consistent with that of main-sequece star forming galaxies \citep[e.g.,][]{Hatcher2021, Mountrichas2022a, Mountrichas2022b}. H$\delta$ and D$_n$4000 indices have been used in the literature to trace recent (few hundred Myrs) star formation bursts in galaxies and as proxies of the age of stellar populations \citep[e.g.,][]{Worthey_ottaviani1997, Kauffmann2003, Wu2018, Sobral2022}. Thus, in this section we compare the two indices between X-ray AGN and sources in the reference galaxy catalogue. 


To facilitate a fair comparison, we need to account for the redshift and stellar mass of the sources in the AGN and reference catalogues. As presented in Fig. \ref{fig_redz}, AGN and galaxies span a narrow redshift range and most importantly have very similar redshift distributions. Fig. \ref{fig_mstar} presents the stellar mass distributions of AGN and sources in the galaxy reference sample. We also plot the M$_*$ distribution of the AGN sample used in \cite{Mountrichas2022a} (see next section). As expected AGN tend to reside in more massive galaxies \citep[e.g.,][]{Yang2017,Yang2018}. To account for this difference  a weight is assigned to each source. This weight is calculated by measuring the joint stellar mass distributions of the two populations (that is, we add the number of AGN and galaxies in each M$_*$ bin, in bins of 0.1\,dex) and then normalise the M$_*$ distributions  by the total number of sources in each bin \citep[e.g.,][]{Mountrichas2019, Masoura2021, Mountrichas2021b, Buat2021}. We make use of these weights in all distributions presented in the remaining of this work.


Fig. \ref{fig_hd_vs_d4000} presents H$_\delta$ vs. D$_n$4000 for AGN and sources in the reference catalogue. Sources located in the upper left corner of the  H$_\delta$-D$_n$4000 space (high H$_\delta$ and low D$_n$4000 values) have predominantly young stellar populations, whereas sources located in the bottom right panel have old stellar populations. We note that the two populations have H$\delta$ and D$_n$4000 with similar uncertainties (see Sect. \ref{sec_legac_catalogue}). We find that $\sim 10\%$ of the AGN are located in the bottom, left corner of the H$_\delta$-D$_n$4000 space, that suggests that although these systems have not undergone a (recent) star formation burst, their stellar population is young. CIGALE's measurements also corroborate to this (f$_{burst}<0.002$ and mean stellar age $\sim 3200$\,Myr). Examination of the optical spectra of these AGN does not show presence of broad emission lines that could contaminate the two indices. We also plot the weighted distributions, i.e., taking into account the different stellar mass distributions of the two populations. The distribution of D$_n$4000 for sources in the reference catalogue appears double peaked, with one peak at D$_n$4000$\sim 1.2$ and a second peak at D$_n$4000$\sim 1.8$. On the other hand, AGN present a peak at  D$_n$4000$\sim 1.4$ and a long tail that extends out to D$_n$4000$\sim 1.8$. Based on the SED fitting calculations presented in the previous section, we estimate the stellar ages that correspond to these D$_n$4000 values. D$_n4000\sim 1.2$ roughly corresponds to 3200\,Myr, while D$_n4000\sim 1.8$ corresponds to $\sim 4700$\,Myr. For the AGN, D$_n4000\sim 1.4$ corresponds to $\sim 3800$\,Myr. A Kolmogorov-Smirnov test (KS-test) gives a p-value of 0.014. This means that the two populations have different D$_n$values at a statistical significant level that is marginally higher than 2\,$\sigma$ \citep[$2\sigma$, which corresponds to a p-value of 0.05, is the minimum threshold to consider two quantities as different, e.g.,][]{Zou2019}. KS-tests are more suitable in finding shifts in probability distributions and present the highest sensitivity around the median value. However, they are less sensitive to the differences at the tails of the distributions, which seem to exist in our case. For that reason, we also apply a Kuiper test that is better at finding spreads that affect the tails of the distributions. The latter gives p-value$=0.002$, i.e., the two populations have different D$_n$4000 distributions at a statistical significant level of $\sim 3\,\sigma$. Regarding the distributions of H$_\delta$, we notice that AGN present a peak at H$_\delta \sim3$ (stellar ages $\sim 4.0$\,Gyr), whereas sources in the reference catalogue have a flatter distribution. KS-test gives p-value$=0.0027$ and Kuiper-test yields p-value$=8.59\times 10^{-9}$ indicating a statistically significant difference in the H$_\delta$ values of AGN and non-AGN systems. \cite{HernanCaballero2014} used 53 X-ray selected AGN at $\rm 0.34<z<1.07$ from the Survey for High-z Absorption Red and Dead Sources (SHARDS) and found a significant excess of AGN host galaxyes with D$_n$4000$\sim 1.4$. Our results are in agreement with theirs.

Overall, based on our analysis, low to moderate luminosity AGN tend to have intermediate D$_n$4000 values, whereas sources in the galaxy reference catalogue prefer low and high D$_n$4000. Moreover, a significant fraction of AGN ($27/69\approx 40\%$) have H$_\delta \sim3$ while non-AGN systems present a flat H$\delta$ distribution. These differences are statistically significant, as indicated by the Kuiper-tests.

\subsection{Spectral indices and position on the main sequence}

In this section, we classify the X-ray AGN and galaxies in the reference catalogue into star forming (SF), starburst (SB) and quiescent (Q) systems and examine how they are distributed in the H$_\delta$-D$_n$4000 space. 

In \cite{Mountrichas2022a, Mountrichas2022b} they classified galaxies as quiescent using their sSFR. Specifically, to identify such systems, they used the location of the second lower peak presented in the sSFR ($\rm sSFR=\frac{SFR}{M_*}$) distributions \citep[section 3.5 of][]{Mountrichas2022a}. We note that in \cite{Mountrichas2022a}, the SFR and M$_*$ of the sources have been calculated using CIGALE with the same templates and parameter values as those used in this study. Using this criterion, we find 42 ($\sim 60\%$) AGN hosted by Q systems in our sample. A similar fraction of Q is found among galaxies in the reference catalogue ($\sim 50\%$). SB systems are identified as those galaxies that have sSFR that is 0.6\,dex higher than the mean value of the \cite{Mountrichas2022a} sample. at the redshift range of our dataset \citep[e.g.][]{Rodighiero2015}. $\sim 8\%$  and $\sim 10\%$ of the AGN and galaxies in the reference catalogue, respectively, are SB. The remaining of the sources that are not Q or SB are considered SF.

The fraction of Q systems appears high compared to the $\sim 25\%$ found in \cite{Mountrichas2022a} at a similar redshift range (see their Table 2). Fig. \ref{fig_mstar} presents the M$_*$ distribution of the X-ray AGN used in \cite{Mountrichas2022a}. We notice that our X-ray sources, that are a subset of the sources used in \cite{Mountrichas2022a}, include the most massive systems of those among the \cite{Mountrichas2022a} sample. This is expected taking into account the selection function of the LEGA-C survey \citep[see Figure A1 in ][]{Wel2021}. This also explains the high fraction of quiescent sources among galaxies in the reference catalogue ($\sim 50\%$ are identified as quiescent).

In Fig. \ref{fig_hd_vs_d4000_ms}, we plot the distribution of the three (host) galaxy classifications in the H$_\delta$-D$_n$4000 space. Quiescent galaxies from the reference catalogue are well separated from the SF and SB non-AGN systems. SB galaxies present the lowest D$_n$4000 and the highest H$\delta$ values. In the case of AGN, SB systems are located in a similar locus with the SB non-AGN galaxies. This is also true for the SF AGN that reside in the same H$_\delta$-D$_n$4000 space with the SF galaxies from the reference sample. In the case of Q AGN, there is a small fraction ($\sim 5\%$) that presents high H$\delta$ and low D$_n$4000, that is, they are located in the H$_\delta$-D$_n$4000 space occupied by SB systems. A similar fraction of Q AGN is also in the H$_\delta$-D$_n$4000 area where SF galaxies are placed. However, the vast majority ($>80\%$) of Q AGN are in the bottom right corner of the plot, where systems with old stellar populations are located.

\subsection{Exclusion of quiescent systems}
\label{sec_quiesc}

In \cite{Mountrichas2022a, Mountrichas2022b} the lower SFR of AGN compared to non-AGN systems was found after excluding quiescent systems from their datasets. To examine whether our results from the comparison of the two indices (H$_\delta$, D$_n$4000) for the two populations corroborate their findings, we exclude from our samples sources that are identified as quiescent (see previous section).



Fig. \ref{fig_hd_vs_d4000_excl_quiescent} presents H$_\delta$ vs. D$_n$4000 for AGN and sources in the reference catalogue, after excluding quiescent systems from both datasets. We notice that for both populations the vast majority of sources (100\% of AGN and 98\% of sources in the reference galaxy catalogue) with high D$_n$4000 values (D$_n$4000$\,>1.6$, which corresponds to about $>4.2$\,Gyr) has been excluded. There are many AGN $(\sim 33\%)$ compared to non-AGN systems ($\sim 7\%$) located within the parameter space defined within $1.2<\rm D_n4000<1.5$ and H$_\delta<3$, that is where sources with old stellar populations are located. Regarding the D$_n$4000 index, both populations peak at similar values ($\rm D_n4000 \sim 1.3$), but sources in the reference catalogue present a wider distribution, as they extend to lower (D$_n$4000$<1.2$) and higher (D$_n$4000$\,>1.55$) values compared to the AGN (p-value$=0.008$ from Kuiper-test). In the case of the  H$_\delta$ distributions, the exclusion of quiescent systems results in the exclusion of the majority of sources with  H$_\delta<2$. However, there is still a significant fraction of AGN with H$_\delta \sim 2-4$. On the other side of the distribution, $\sim 20\%$ of the galaxies have H$_\delta>6$, whereas the corresponding fraction of AGN is $\sim 5\%$. Although, these differences do not appear to be statistically significant (p-value$=0.08$, $<2\,\sigma$, based on Kuiper-test), they suggest that AGN that reside in star forming (non-quiescent) galaxies tend to have on average lower H$_\delta$ values compared to their non-AGN star forming counterparts. 

We also identify and exclude sources as quiescent, using their D$_n$4000 and H$_\delta$ values. Specifically, we classify as quiescent systems that have D$_n$4000$>1.55$ and/or EW(H$_\delta)<2\AA$ \citep{Wu2018}. This analysis is presented in Appendix \ref{sec_appendix}. The results and conclusions are similar to those presented above.

Based on our analysis, when quiescent systems are excluded, low to moderate luminosity AGN tend to live in systems with older stellar populations and are less likely to have experienced a recent burst of their star formation compared to their non-AGN counterparts. The results are not sensitive to the method applied to select quiescent systems. This could explain the results presented in \cite{Mountrichas2022a, Mountrichas2022b}, where they found that low to moderate luminosity (non-quiescent) AGN have slightly (by $\sim20\%$) lower SFR compared to SF galaxies.

\subsection{The role of (host-)galaxy shape}

\cite{Yang2019} used a sample from the five CANDELS fields \citep[][]{Grogin2011, Koekemoer2011} and found that the supermassive black hole (SMBH) accretion rate (BHAR) correlates stronger with SFR than with M$_*$, for bulge-dominated (BD) systems. The term BD refers to galaxies that only display a significant spheroidal component, without obvious discy or irregular components \citep[for more details see Sect. 2.3 in][]{Yang2019}. \cite{Ni2021} used sources from the COSMOS field and found that for star forming BD and non-BD systems, BHAR correlates more with $\Sigma _1$ than with SFR or M$_*$, where $\Sigma _1$ measures the mass-to-size ratio in the central 1\,kpc of galaxies and is used as a proxy of a galaxy compactness. 

In this section, we use the catalogue presented in \cite{Ni2021} and cross-correlate it with our AGN and galaxy reference catalogues to add information about the morphology and compactness of our systems. The morphological classification has been made using a deep-learning-based method to separate sources as BD and non-BD galaxies \citep[for more information see Appendix C in][]{Ni2021}.  $\Sigma _1$ has been measured by fitting sources with a single component S\'{e}rsic profile and assuming a constant M$_*$-to-light ratio throughout the galaxy \citep[for a detailed description see Sect. 2.2 in][]{Ni2021}. We note that type 1 sources have been removed from the \cite{Ni2021} catalogue, since they can potentially affect host galaxy morphological measurements (see their section 2.4). 

From the 69 AGN and 2176 sources in the reference catalogue used in our analysis above, 57 and 1782, respectively, are included in the \cite{Ni2021} dataset. Out of the 57 AGN, 16 ($28\%$) are classified as BD and a similar fraction is found among galaxies in the reference catalogue (459 sources - $25\%$). 

In Fig. \ref{d4000_hdelta_bd}, we plot the D$_n$4000, H$\delta$ distributions of BD and non-BD AGN and galaxies. Distributions are weighted based on the M$_*$ of the sources. The D$_n$4000 distribution of galaxies classified as BD peaks at higher values compared to non-BD galaxies, which implies that BD galaxies tend to have older stellar populations compared to non-BD systems. This is in agreement with previous studies that have found that the fraction of BD drops significantly towards high SFR \citep[e.g.,][morphological quenching]{Huertas2016, Martig2009}. Comparing AGN with non-AGN systems, the D$_n$4000 distribution of non-BD AGN peaks at higher values compared to non-BD galaxies, while the opposite is true when comparing BD AGN with BD, non-AGN galaxies.

Regarding the H$\delta$ values, the distribution of non-BD galaxies peaks at high H$_\delta$ indicating that most of these systems have undergone a recent burst of their star formation. The H$_\delta$ distribution of BD galaxies peaks at significantly lower values, consistent with the picture that these systems include old(er) stellar populations. A comparison between non-BD systems shows that AGN tend to have lower H$_\delta$ values compared to non-AGN galaxies. BD AGN have a flat H$_\delta$ distribution, whereas BD non-AGN galaxies peak at low H$\delta$ values.





In Fig. \ref{d4000_hdelta_S1_both}, we plot D$_n$4000 (top panel) and H$_\delta$ (bottom panel) as a function of the compactness, $\Sigma _1$ (mass-to-size ratio), for AGN and galaxies. An immediate difference to notice is that AGN do not have $\Sigma _1$ values lower than 9.0, whereas sources in the reference catalogue go down to 7.5. This implies that AGN reside in more compact systems compared to non-AGN. Both populations present an increase of D$_n$4000 and a decrease of H$_\delta$ with $\Sigma _1$, at $\Sigma _1 \geqslant 9.5$. Interestingly, this $\Sigma _1$ value is similar to the median $\Sigma _1$ values of the two samples (9.8 for the AGN and 9.6 for the galaxies). For both populations, we find that the vast majority ($>95\%$) of BD systems have  $\Sigma _1 > 9.5$.

Based on our results, the morphology of a galaxy seems to affect more the stellar population of non-AGN systems as opposed to AGN, since galaxies that host AGN appear to have stellar populations with similar age and are equally likely to have experienced a recent burst, regardless of their morphological type. BD AGN tend to have younger stellar populations compared to BD non-AGN systems. On the other hand, non-BD AGN have, on average, older stellar populations compared to non-BD sources in the reference catalogue. Finally, AGN prefer more compact systems compared to non-AGN, based on the mass-to-size ratio.

\section{Summary-Conclusions}

We used 69, low to moderate L$_X$ ($\rm L_{X,2-10keV} \sim 10^{42.5-44}\,erg\,s^{-1}$), X-ray AGN from the UltraVISTA region of the $\it{COSMOS-Legacy}$ field that are also included in the LEGA-C catalogue. The latter, provides measurements on absorption line indices, among others. We also constructed a galaxy reference catalogue that consists of 2176 sources. The two populations have the same photometric coverage and lie within $\rm 0.6<z<1.3$. We make use of two spectral indices, provided by the LEGA-C catalogue, namely the D$_n$4000 and the H$_\delta$, that are sensitive to stellar ages. D$_n$4000 increases monotonically with time while the peak strength of H$_\delta$ depends on whether the SFR varies rapidly or changes smoothly.

Our purpose is to examine if the inclusion of the two indices provides additional constraints on the SED fitting process and thus affects the (host) galaxy properties. Moreover, we compare the values of the two indices between AGN and non-AGN systems (reference catalogue) to extract information regarding their stellar populations.

To examine if the inclusion of D$_n$4000 and H$_\delta$ affects the SFH parameters and important (host) galaxy properties (SFR, M$_*$), we use the CIGALE SED fitting code. Our analysis revealed that adding the two indices in the fitting process allows CIGALE to constrain better the stellar ages and the e-folding time of the stellar population. We also found an increase of the M$_*$ measurements by $\sim 0.2$\,dex that is more evident for systems that host young stellar populations (D$_n$4000$<1.5$). The results are similar for AGN and sources in the reference catalogue. We found that these changes should  mostly be attributed to the addition of the H$_\delta$ line rather than the D$_n$4000 index. The stability we found on global SFH parameters (SFR, M$_*$) is attributed to the very good photometric coverage of our sample. In a future work, we will examine the effect of adding more spectroscopic information in the SED fitting and in relation to the photometric coverage. 

We then compared the D$_n$4000 and H$_\delta$ for AGN and sources in the reference catalogue. The two populations have similar redshift distributions, therefore for the comparison we accounted only for their different M$_*$, by assigning weighs to the sources in the two datasets. Our analysis reveals that AGN tend to reside in systems with intermediate stellar ages (D$_n$4000$\sim 1.4$), whereas sources in the reference catalogue presented a double-peaked D$_n$4000 distribution. The latter also has a flat H$_\delta$ distribution while AGN peaks at H$_\delta \sim 3\AA$. These differences are statistically significant (p-values$=0.002$ and $8.59\times 10^{-9}$, for D$_n$4000 and H$_\delta$, respectively). 

When we excluded quiescent systems from both populations, we found that the low to moderate X-ray AGN used in our analysis tend to live in systems with older stellar population (higher D$_n$ values) and are less likely to have experienced a recent burst (lower H$_\delta$ values), compared to galaxies in the reference catalogue. This is in agreement with the results presented in \cite{Mountrichas2022a, Mountrichas2022b}, where they found that AGN with similar luminosities with those used in this work, have lower SFR (by $\sim 20\%$) compared to SF galaxies. 

We, also, compared the two indices for AGN and galaxies with different morphologies (BD and non-BD) as well as based on their mass-to-size ratio ($\Sigma _1$). A similar fraction of AGN and non-AGN systems are classified as non-BD ($\sim 70\%$). Our analysis showed that BD AGN tend to have younger stellar populations (lower D$_n$4000) compared to BD non-AGN systems. On the other hand, non-BD AGN have, on average, older stellar populations (higher D$_n$4000) and are less likely to have experienced a recent burst (lower H$\delta$ values) compared to non-BD sources in the reference catalogue. Based on our analysis, the morphology of a galaxy seems to affect more the stellar population of non-AGN systems as opposed to AGN, since galaxies that host AGN appear to have stellar populations with similar age and are almost, equally likely to have experienced a recent burst, regardless of their morphological type. Furthermore, AGN prefer more compact systems compared to non-AGN, based on the mass-to-size ratio. 

Our work showed that low to moderate X-ray luminosity AGN and non-AGN galaxies are systems that, on average, have different stellar ages and different likelihood of having undergone a recent SF. Different trends are also found for the two populations based on the shape of the (host) galaxy.

\begin{acknowledgements}
We thank the referee for their careful reading of the paper. GM acknowledges support by the Agencia Estatal de Investigación, Unidad de Excelencia María de Maeztu, ref. MDM-2017-0765.
The project has received funding from Excellence Initiative of Aix-Marseille University - AMIDEX, a French 'Investissements d'Avenir' programme.
This work was partially funded by the ANID BASAL project FB210003. MB acknowledges support from FONDECYT regular grant 1211000.
QN acknowledges support from a UKRI Future
Leaders Fellowship (grant code: MR/T020989/1)

\end{acknowledgements}

\bibliography{mybib}{}
\bibliographystyle{aa}



\appendix

\section{Exclusion of quiescent systems based on their D$_n$4000 and H$\delta$ values}
\label{sec_appendix}

\begin{figure}
\centering
  \includegraphics[width=0.9\columnwidth, height=6.5cm]{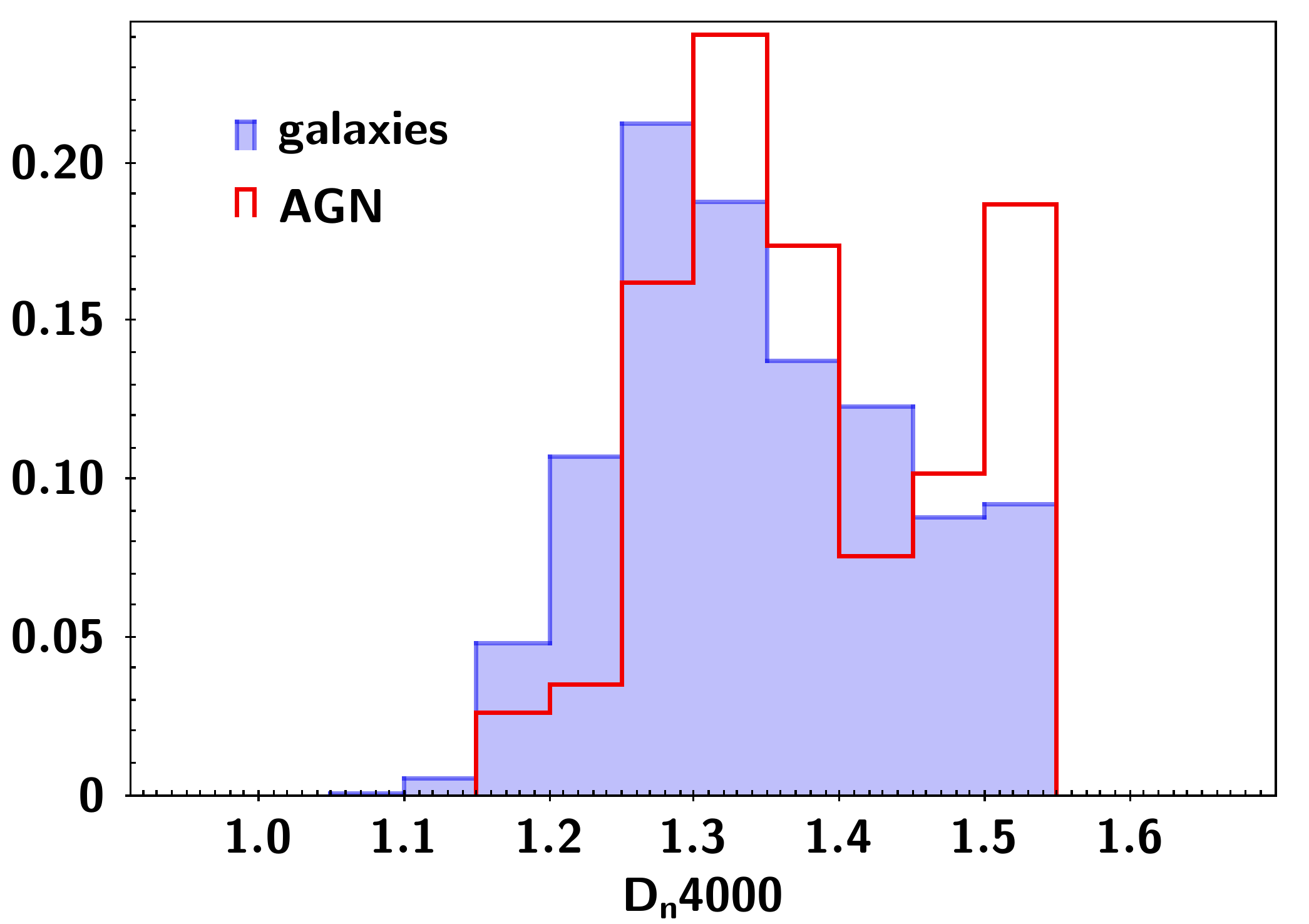} 
  \includegraphics[width=0.9\columnwidth, height=6.5cm]{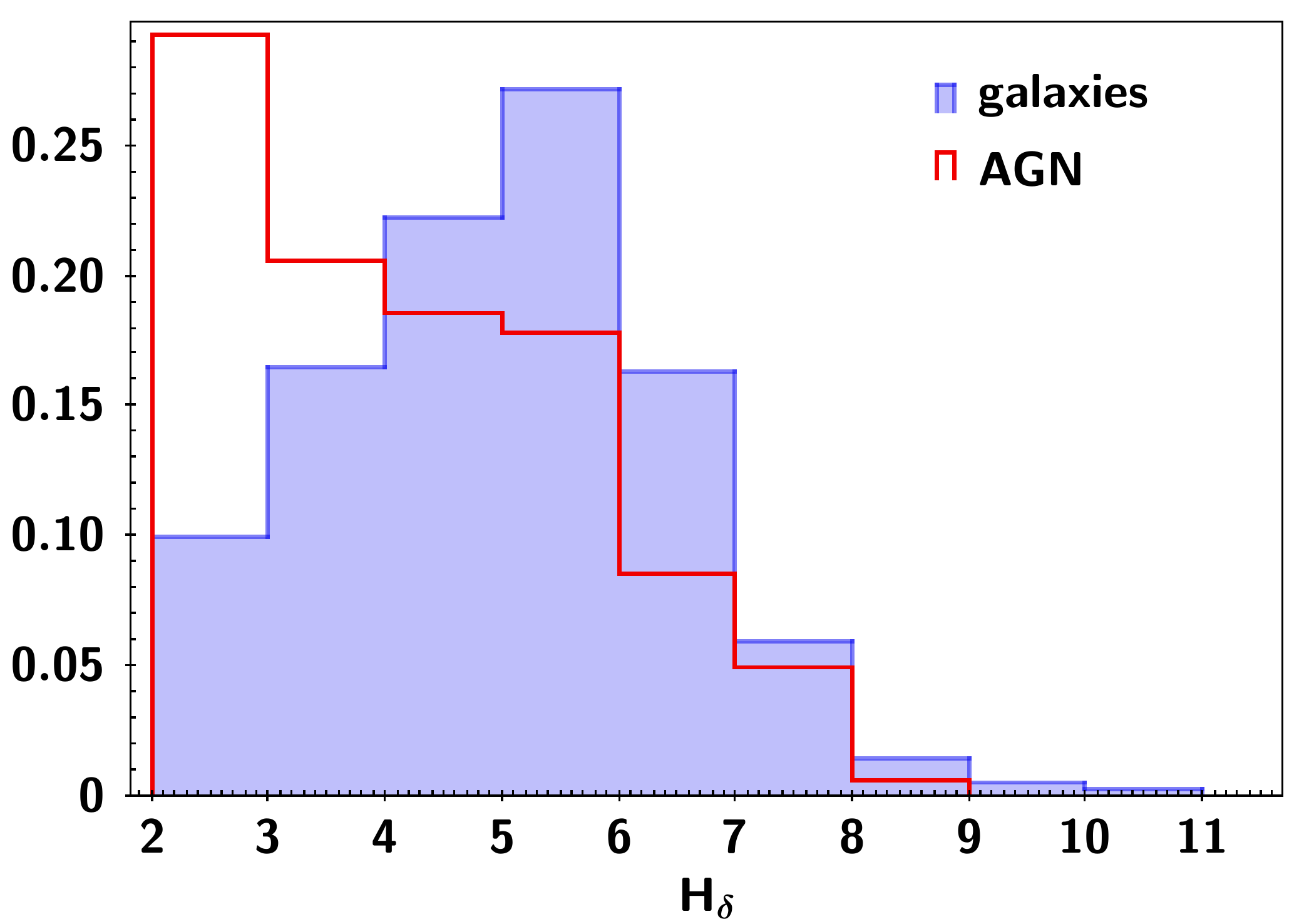} 
  \caption{Comparison of the distributions of D$_n$4000 and H$_\delta$ (in \AA) for galaxies in the reference catalogue and AGN, when excluding systems as quiescent systems based on D$_n$4000 and H$_\delta$ (see text for more details). Sources in the reference catalogue tend to have younger stellar population (smaller D$_n$4000) and are more likely to have undergone a recent star formation burst (higher H$_\delta$), compared to AGN.} 
  \label{fig_d4000_new_quiesc}
\end{figure}

\cite{Wu2018} used galaxies from the LEGA-C dataset and examined their stellar ages and SFH histories using D$_n$4000 and H$_\delta$. In their analysis, they identify quiescent systems utilising the two indices. Specifically, they classify as quiescent systems that have D$_n$4000$\,>1.55$ and/or EW(H$_\delta)<2\AA$. We apply their criteria on our AGN and reference catalogue to check whether the results presented above are sensitive to the choice of selecting quiescent systems. Based on these criteria, there are 51 non-quiescent AGN in the X-ray sample and 1401 non-quiescent sources in the galaxy reference catalogue. We then compare the D$_n$4000 and H$_\delta$ distributions of the two populations, in Fig. \ref{fig_d4000_new_quiesc}. The results and conclusions are similar to those presented in Sect. \ref{sec_quiesc}, where we identified quiescent systems based on their sSFR. Specifically, non-quiescent AGN present a large(r) tail that extends to high D$_n$4000 values compared to non-quiescent systems in the reference catalogue (top panel of Fig. \ref{fig_d4000_new_quiesc}). The H$\delta$ distributions of the two populations also appear different (bottom panel of Fig. \ref{fig_d4000_new_quiesc}), with AGN to peak at H$\delta = 2-3\,\AA$ and sources in the reference catalogue at H$\delta = 5-6\,\AA$.

\end{document}